\documentclass[number,sort&compress,twocolumn,5p]{elsarticle}
\usepackage[binary-units=true,separate-uncertainty=true]{siunitx}
\usepackage{lineno,hyperref}
\usepackage{tikz}
\usepackage{amssymb,amsmath,mathtools}
\usepackage{caption}
\usepackage{xspace}
\usepackage{booktabs}
\usepackage{siunitx}
\usepackage{textcomp}

\usepackage[percent]{overpic}
\usepackage[list=true,listformat=simple,margin=10pt]{subcaption}
\usepackage{hepnames}

\journal{Nucl. Instr. Meth. A}

\begin{document}
\begin{frontmatter}

\title{Comparison of different sensor thicknesses and substrate materials for the monolithic small collection-electrode technology demonstrator CLICTD}

\author[cern]{K.~Dort\corref{corr}\fnref{ugiessen}}
\ead{katharina.dort@cern.ch}
\author[cern]{R.~Ballabriga}
\author[cern]{J.~Braach\fnref{uhamburg}}
\author[cern]{E.~Buschmann}
\author[cern]{M.~Campbell}
\author[cern]{D.~Dannheim}
\author[desy]{L. Huth}
\author[cern]{I.~Kremastiotis}
\author[cern]{J.~Kr\"oger\fnref{heidelberg}}
\author[cern]{L.~Linssen}
\author[ugeneva]{M.~Munker}
\author[cern]{W.~Snoeys}
\author[desy]{S.~Spannagel}
\author[cern]{P.~\v{S}vihra}
\author[desy]{T.~Vanat}

\address[cern]{CERN, Geneva, Switzerland}
\address[desy]{Deutsches Elektronen-Synchrotron DESY, Notkestr. 85, 22607 Hamburg, Germany}
\address[ugeneva]{University of Geneva, Geneva, Switzerland}

\cortext[corr]{Corresponding author}

\fntext[uhamburg]{Also at University of Hamburg, Germany}
\fntext[ugiessen]{Also at University of Giessen, Germany}
\fntext[heidelberg]{Also at University of Heidelberg, Germany}

\begin{abstract}
   Small collection-electrode monolithic CMOS sensors profit from a high signal-to-noise ratio and a small power consumption, but have a limited active sensor volume due to the fabrication process based on thin high-resistivity epitaxial layers. 
   In this paper, the active sensor depth is investigated in the monolithic small collection-electrode technology demonstrator CLICTD. 
   Charged particle beams are used to study the charge-collection properties and the performance of devices with different thicknesses both for perpendicular and inclined particle incidence. 
   In CMOS sensors with a high-resistivity Czochralski substrate, the depth of the sensitive volume is found to increase by a factor two in comparison with standard epitaxial material and leads to significant improvements in the hit-detection efficiency and the spatial and time resolution.

\end{abstract}


\end{frontmatter}


\section{Introduction}
\label{sec:introduction}
In monolithic CMOS sensors the readout electronics are integrated into the active sensor volume, which offers the potential for fine pixel pitches and a low mass.
Profiting from the commercial CMOS industry, these devices are particularly suited for large-scale production.
Monolithic sensor designs featuring a small collection electrode benefit from a reduced  capacitance, which enables an improvement in signal-to-noise ratio and reduced power consumption~\cite{Snoeys}.
Several sensors have been fabricated in a modified 180\,nm CMOS imaging process implementing the small collection-electrode design with a 25 - \SI{30}{\micro \meter} high-resistivity epitaxial layer on a low-resistivity substrate, such as the ALPIDE~\cite{mager2016alpide}, (Mini-)MALTA~\cite{dyndal2020mini}, FASTPix~\cite{kugathasan2020monolithic} and CLICTD~\cite{clictdTestbeam} sensors. 
They have exhibited promising results regarding radiation tolerance, a time resolution down to hundreds of picoseconds, a spatial resolution of a few micrometers and full efficiency over a wide threshold range. 

Although sensor optimisations enable a full lateral depletion~\cite{tj-modified} in the small collection-electrode design, the devices are only partially depleted in depth. 
The active sensor depth, from which charge carriers contribute to the signal, extends further than the depletion depth but is limited by the thickness of the epitaxial layer due to the short charge carrier lifetime in the low-resistivity substrate.  
High-resistivity substrate materials are therefore investigated as a possible replacement, extending both the depletion and active sensor depth thus leading to a higher measured signal.
In this document, a high-resistivity Czochralski substrate as alternative wafer material is assessed, which has already proven to increase efficiency after irradiation in the small collection-electrode design~\cite{pernegger2021radiation}. 

An in-depth comparison of 40 - \SI{300}{\micro \meter} thick sensors in the original epitaxial-layer design with \SI{100}{\micro \meter} thick Czochralski sensors is presented for the CLICTD technology demonstrator. 
To this end, the performance and charge-sharing characteristics of different CLICTD sensors are studied using charged particle beams with perpendicular and inclined incidence relative to the sensor surface. 
Most notably, in-pixel studies are presented that allow for a two-dimensional mapping of charge-collection properties. 
The effective active sensor volume is determined as well by employing the grazing angle technique~\cite{meroli2011grazing} for the different sensor thicknesses and materials.


\section{The CLICTD Sensor}
\label{sec:clictd}

\begin{figure*}[bpt]
	\centering
	\begin{subfigure}[t]{0.49\textwidth}
		\centering
		\includegraphics[width=0.85\columnwidth]{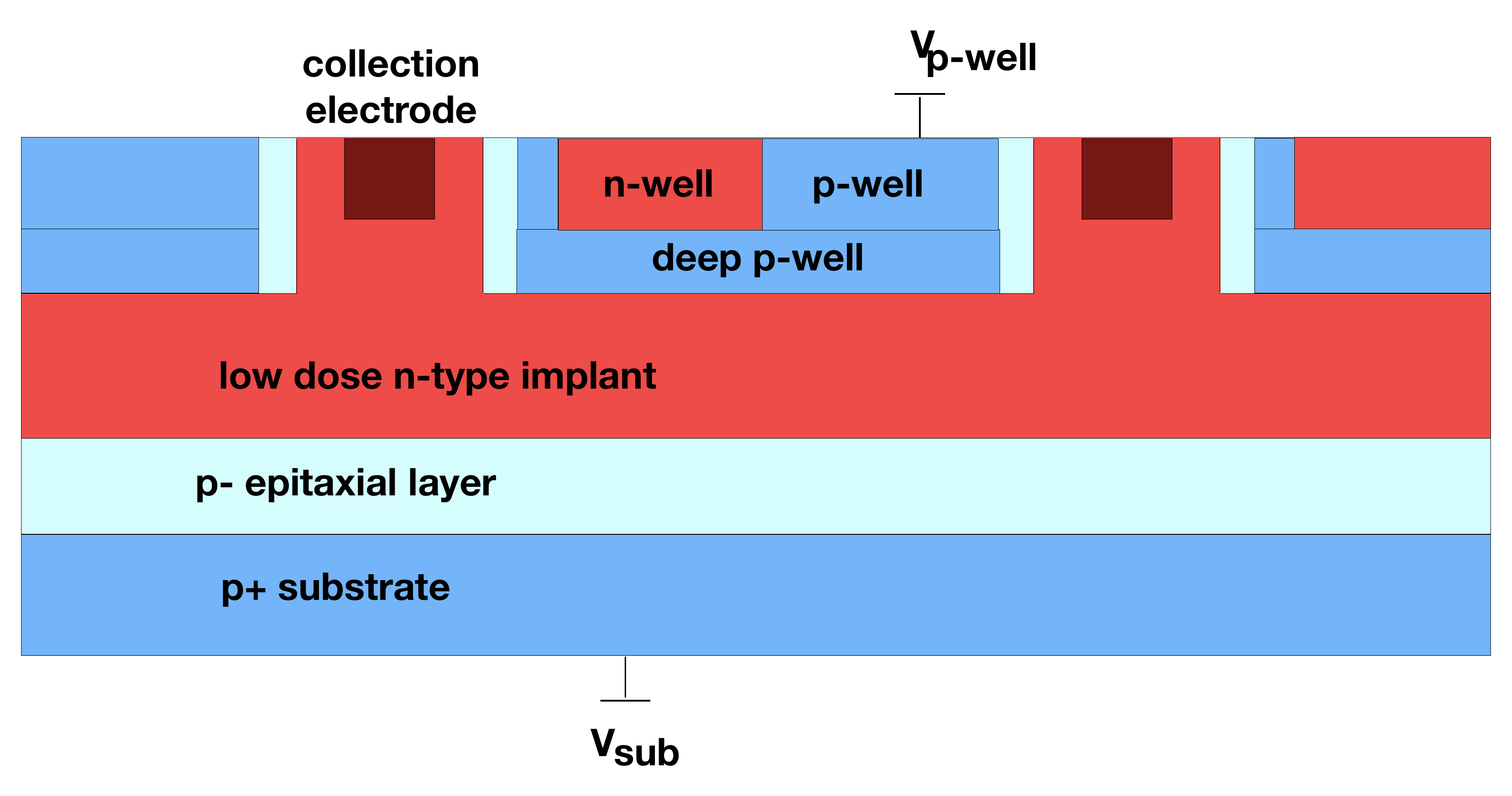}%
		\caption{Continuous n-implant}
		\label{fig:mod_flavour}
	\end{subfigure}
	\begin{subfigure}[t]{0.49\textwidth}
		\centering
		\includegraphics[width=0.85\columnwidth]{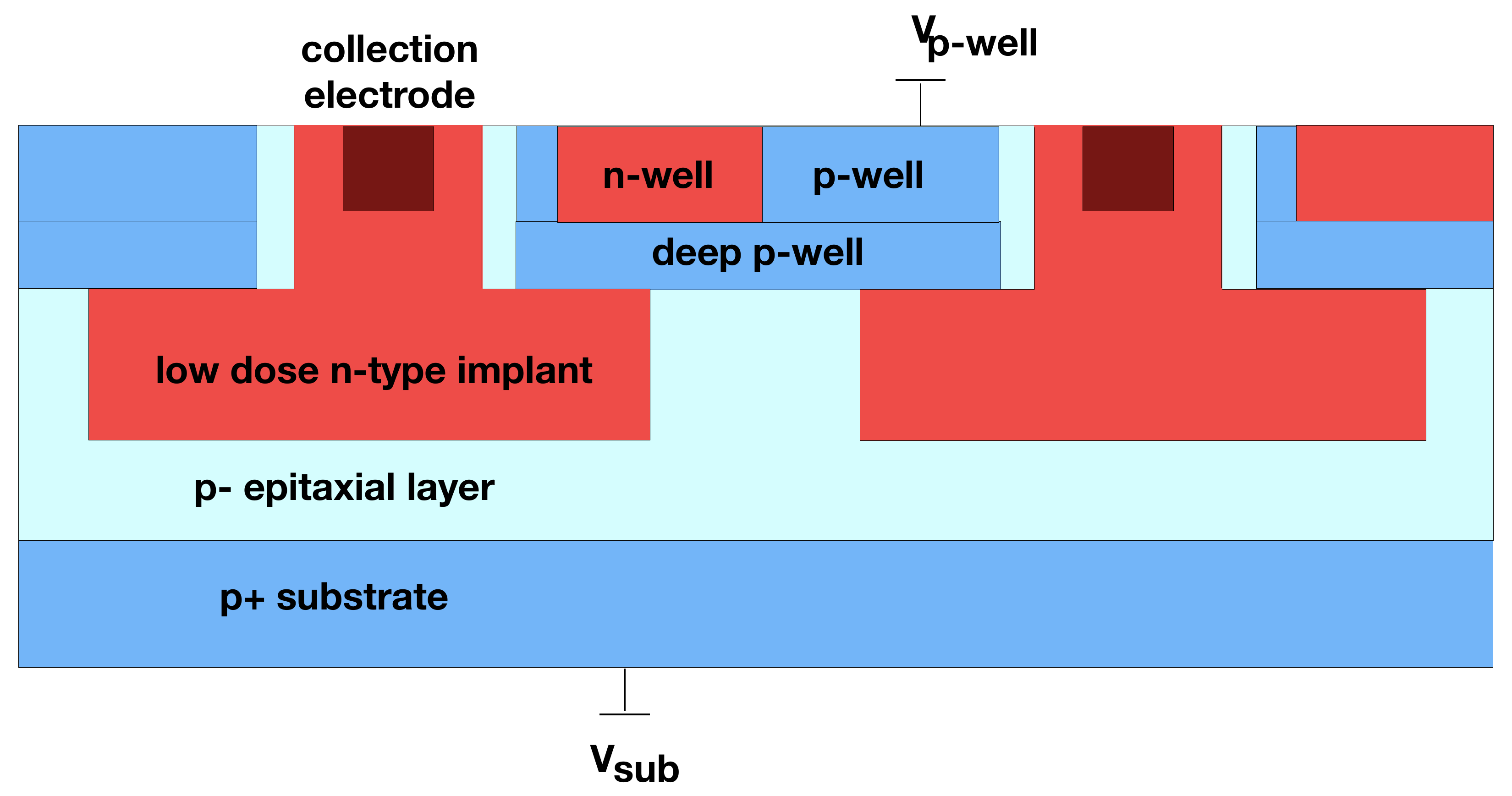}%
		\caption{Segmented n-implant}
		\label{fig:gap_flavour}
	\end{subfigure}%
	\caption{Schematics of the CLICTD pixel design for the pixel flavour with (a) continuous and (b) segmented n-implant.
	The readout circuitry is placed in the nwell and pwell above the deep pwell.
	Not to scale.}
	\label{fig:clictdFlavours}
\end{figure*}

The CLICTD sensor features a matrix of $16\times128$ detection channels with a size of \SI{300x30}{\micro m} in column\,$\times$\,row direction.
Each channel is segmented along the column direction into 8 sub-pixels with a size of \SI{37.5x30}{\micro m}.
The following section gives a brief overview of the main features of the CLICTD sensor.
Additional details can be found in~\cite{clictdTestbeam} and~\cite{clictd_design_characterization}.

\subsection{Sensor Design}

The CLICTD sensor is fabricated in a modified 180\,nm CMOS imaging process~\cite{tj-modified} using two different pixel flavours, as shown schematically in Fig.~\ref{fig:clictdFlavours}.
The sensor is characterised by a small n-type collection electrode on top of a \SI{30}{\micro \meter} thin high-resistivity (few k$\Omega$cm) p-epitaxial layer, that is grown on a low-resistivity ($\sim 10^{-2}\SI{}{\ohm \centi \metre}$) p-type substrate.
The on-channel front-end electronics is placed in the n-well and p-well above the deep p-well, which shields the circuitry.
A low-dose n-type implant below the p-wells allows for full lateral depletion of the epitaxial layer~\cite{tj-modified}.
In the second pixel flavour, the n-implant is segmented at the pixel edges, which causes an increase in the lateral electric field.
As a consequence, an accelerated charge collection and reduced charge sharing is achieved with this design.  
In the CLICTD sensor, the segmentation is only introduced in the column direction. 
In the row direction, a high degree of charge sharing is desired in order to improve the spatial resolution. 

A reverse bias voltage is applied to nodes in the p-wells and the substrate. 
The bias voltage at the p-wells is limited to -6\,V to prevent breakdown of the on-channel NMOS transistors~\cite{thesis-jacobus}.

\subsection{Sensor Material}

CLICTD sensors with different thicknesses were produced using backside grinding. 
The total device thickness ranges from \SI{40}{\micro \meter} to \SI{300}{\micro \meter}, including a  metal stack of approximately \SI{10}{\micro \meter} on top of the sensor~\cite{prabket2019resistivity}.

The size of the active sensor volume is limited by the thickness of the \SI{30}{\micro \meter} epitaxial layer. 
To increase the active volume, an alternative substrate material is studied, which consists of high-resistivity (few \SI{}{\kilo \ohm \cm}) p-type Czochralski silicon~\cite{pernegger2021radiation}.
The implants are introduced directly on the Czochralski wafers and no additional epitaxial layer is grown on top.
Henceforth, the term \textit{Czochralski substrate} is used to refer to the sensors fabricated on high-resistivity Czochralski wafers.

The advantages of the high-resistivity Czochralski material are twofold: 
Firstly, the isolation between p-well and substrate bias nodes is improved, allowing for a larger difference between the two voltages. 
Secondly, the depletion can evolve further in depth owing to the larger size of the high-resistivity volume. 

The benefits of the larger active volume depend on the aspect ratio of the pixel cell and the target applications of the sensor.
For instance, sensors with a comparably large pixel pitch that aim for a good spatial resolution, profit from a larger active depth by tuning the depth such that an optimal degree of charge sharing and an enhanced signal are achieved. 
Sensors with a pixel pitch considerably smaller than the active depth are less suited for the Czochralski substrate, since the cluster size increases considerably, which typically does not lead to additional improvements of the performance. 
Likewise, in designs where the charge collection within the sensor is subject to a high degree of charge sharing, thick sensors with Czochralski substrate are less suited. 
Therefore,  the Czochralski substrate is only investigated for CLICTD sensors with segmented n-implant, since charge sharing is suppressed for this pixel flavour. 

It should be noted that the availability of high-resistivity Czochralski substrates for silicon sensor fabrication depends on foundry specifications, since it is not a standard material for the investigated CMOS process. 

\subsection{Analogue and Digital Front-End}

Each sub-pixel has an analogue front-end that consists of a voltage amplifier connected to a discriminator, where an adjustable detection threshold is compared to the input pulses.  
Effective threshold variations are corrected using a 3-bit threshold-tuning DAC. 

The discriminator output of the eight sub-pixels in a detection channel are combined with a logical \textit{OR} in the on-channel digital front-end.
The binary hit pattern of the sub-pixels is recorded as well as the 8-bit Time-of-Arrival (ToA) and the 5-bit Time-over-Threshold (ToT) for time and energy measurements, respectively. 
As a consequence of combining the sub-pixel discriminator outputs, the ToA is set by the earliest sub-pixel timestamp and the ToT is determined by the number of clock cycles in which at least one sub-pixel is above the detection threshold. 

No conversion from ToT to physical units is applied for the measurements shown in this paper, since the conversion was found to have a limited precision owing to non-linearities in the analogue front-end~\cite{clictdTestbeam}.

\subsection{Sensor Operation}

The front-end and operation settings were optimised in laboratory studies detailed elsewhere~\cite{clictd_design_characterization, clictdTestbeam}.
Most importantly, for each sensor a minimum operation threshold is defined as the lowest possible threshold at which a noise free operation ($< 1 \times 10^{-3}$\,hits/s for the full pixel matrix) is achievable with up to 10 noisy pixels masked, which is less than one per mille of the entire matrix. 
The sensors presented in this paper are compared at their respective minimum operation threshold. 
It should be noted that measurements below the minimum operation threshold are nevertheless feasible, since a small noise contribution can be tolerated. 

The difference between the substrate and p-well bias voltages is limited by the punch-through between the two nodes. 
Whereas this requirement constraints the difference to a few volts for sensors with epitaxial layer, for the Czochralski sensors, the difference can easily exceed tens of volts. 
For the sensors with epitaxial layer, a high substrate bias voltage has a negligible impact, since the depletion depth is limited by the thickness of the epitaxial layer itself.
Therefore, the bias voltage is fixed to -6\,V/-6\,V at the p-well/substrate nodes for measurements presented in the following sections. 
For the Czochralski sensors, the depletion region can evolve further into the substrate, thus justifying measurements with increased substrate bias voltage.

\subsection{Front-End Optimisation for Large Substrate Voltages}

The CLICTD front-end is optimised for sensors with a \SI{30}{\micro \meter} epitaxial layer. 
Sensors fabricated on  Czochralski substrates are subject to a higher sensor leakage current, if the difference between p-well and substrate voltage exceeds 5\,V. 
The increased current can saturate the leakage current compensation circuit, which renders parts of the pixel matrix insensitive to incoming particles. 
To counteract the saturation, the front-end settings are adapted such that a faster return to baseline at the input node is achieved. 
With these settings, the sensor can be operated up to -20\,V substrate and -6\,V p-well bias voltage before any saturation effects set in. 
However, the adaptations reduce the signal gain, which leads to coarser steps in the threshold settings and a larger minimum operation threshold, since the front-end is operated in conditions it was not designed for. 
The higher thresholds have important implications for the sensor performance, as presented in Section~\ref{sec:performance}. 

The total power consumption of the device has been studied elsewhere~\cite{clictd_design_characterization}. 
As the adapted settings at the input node only increase the range of the leakage current compensation, the current and power consumption of the main circuit is not affected.


\section{Test-Beam and Analysis Setup}
\label{sec:setup}

\begin{figure*}[tbp]
	\centering
	\includegraphics[width=0.7\textwidth]{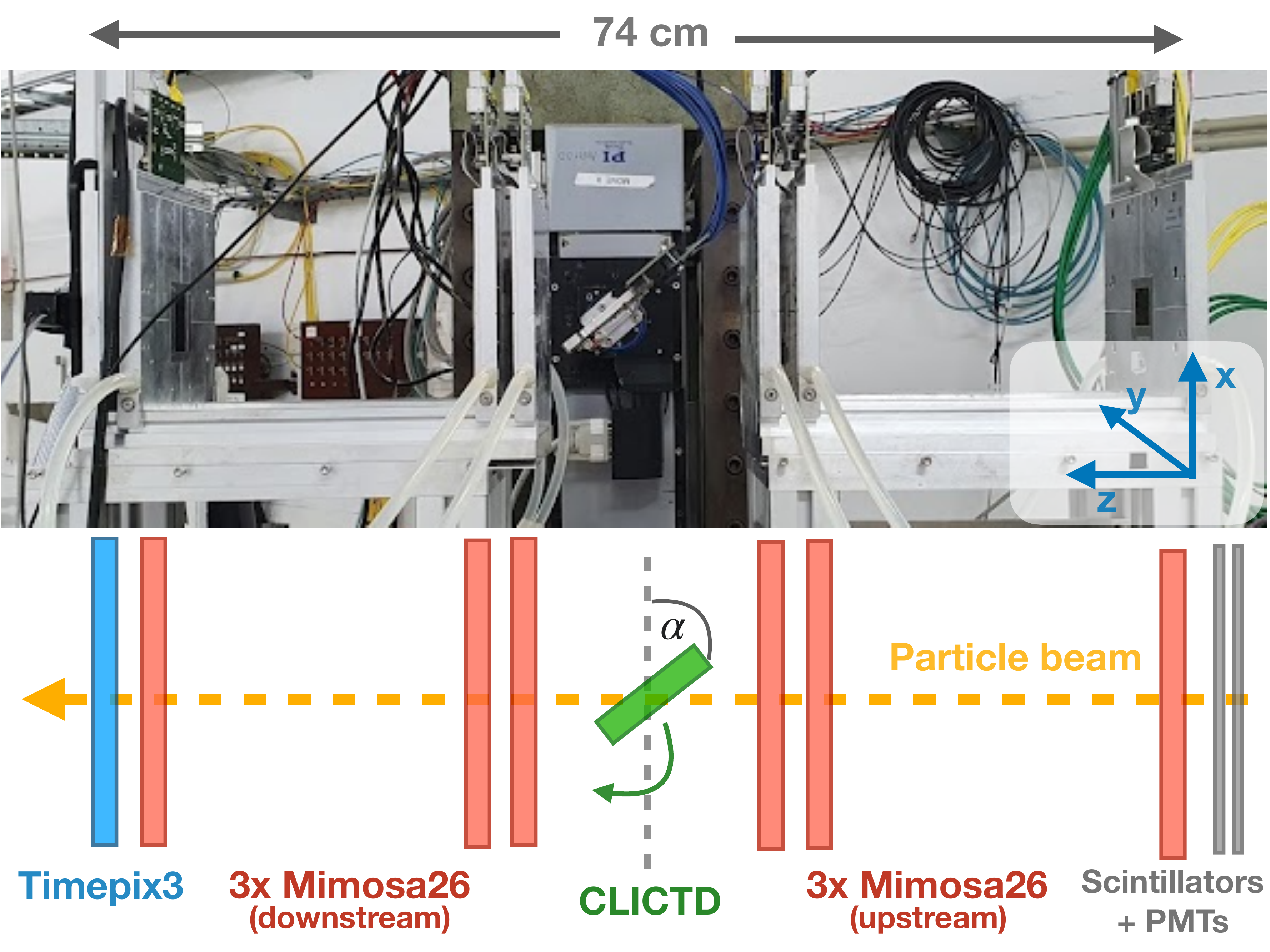}%
	\caption{Test-beam setup with a rotated DUT}
	\label{fig:testbeam-setup}
\end{figure*}

Test-beam measurements were performed in a two-week test-beam period at the DESY\,II Test-Beam Facility~\cite{Diener:2018qap} using a MIMOSA-26 telescope~\cite{baudot2009first} equipped with an additional Timepix3~\cite{Poikela_2014} plane for improved track-time resolution, as schematically depicted in Fig.~\ref{fig:testbeam-setup}. 
The beam consisted of 5.4\,GeV electrons and data for different incidence angles between the beam and the sensor surface were recorded. 
To this end, the Device-Under-Test (DUT) was mounted on a rotation stage to allow for inclinations relative to the beam axis. 

Two different telescope plane spacings were used to optimise the tracking performance for the respective measurements: For measurements with perpendicular incidence between the beam and the sensor surface, the innermost telescope planes are as close as physically possible to the DUT.
When the DUT is rotated, the telescope planes are adjusted such that the DUT can be tilted to $\leq70^\circ$ without touching the telescope planes.

A trigger signal, consisting of a coincidence between two scintillators in front of the first telescope plane, is provided by the AIDA Trigger Logic Unit (TLU)~\cite{Baesso:2019smg}. 
The \mbox{EUDAQ2} data acquisition framework is used to control and read out the telescope and the DUT~\cite{Liu:2019wim}.

\subsection{Reconstruction and Analysis}

The software framework Corryvreckan~\cite{corry_paper, corry_manual} is used to perform offline reconstruction and analysis of the test-beam data.
Individual events are defined by CLICTD readout frames.
The start of each frame is marked by opening the acquisition shutter and the end by closing the shutter 200\,ns after receiving a trigger signal from the TLU.
The Timepix3 hit timestamp and the TLU trigger timestamp associated to MIMOSA-26 hits determine their allocation to a specific event by requiring that the timestamp is within a CLICTD frame. 
The subsequent analysis is performed on an event-by-event basis.

For each telescope plane and the DUT, adjacent pixel hits are combined into clusters and the cluster position is calculated by a ToT-weighted centre-of-gravity algorithm.
For the CLICTD sensor, the cluster position in row direction is corrected using the $\eta$-formalism to take non-linear charge sharing between pixel cells into account~\cite{Akiba:2011vn, BELAU1983253, clictdTestbeam}. 
In addition, \textit{split clusters} are considered for measurements with rotated DUT i.e. a gap of one pixel is permitted between pixel hits to account for single pixels within a cluster that fall below the threshold. 
The risk of having \textit{merged clusters} from two distinct particles is considered to be negligible, since less than one percent of all events contain more than one reconstructed track.

Track candidates are formed from clusters on each of the seven telescope planes.
For track fitting the General Broken Lines (GBL) formalism~\cite{Blobel:2006yi} is used to account for multiple scattering in the material. 
The telescope alignment is performed by minimising the track $\chi^2$ distribution. 
Tracks with a $\chi^2$ per degree of freedom larger than three are discarded. 
The telescope track resolution at the position of the DUT is \SI{2.4}{\micro \meter} for the close telescope plane spacing and \SI{5.6}{\micro \meter} for the wide rotation configuration, as estimated from analytical calculations based on~\cite{resolutionSimulator, Jansen:2016bkd}. 

A reconstructed track is associated with a CLICTD cluster by requiring a spatial distance of less than 1.5 pixel pitches between the global track intercept position on the DUT and the reconstructed cluster position as well as a track timestamp within the same CLICTD frame as the cluster.
It has been verified that the spatial cut is sufficiently large even for the larger track resolution at the position of the DUT in the wide telescope-plane configuration.
Clusters adjacent to the edge of the pixel matrix are rejected to exclude edge effects.
The following observables are considered to characterise the DUT:
 
\paragraph{Cluster size}

The cluster size is defined as the number of pixels in a given cluster.
Correspondingly, the cluster size in column/row direction is given by the size of the cluster projected onto the respective axis.
The systematic uncertainty on the cluster size arises from uncertainties in the threshold calibration, as detailed in~\cite{clictdTestbeam}.
At the minimum operation threshold, the systematic uncertainty evaluates to $\pm 0.01$ for the mean cluster size and the statistical uncertainty is of the order of $10^{-4}$.

\paragraph{Hit-detection efficiency}
The hit-detection efficiency is calculated as the number of associated tracks divided by the total number of tracks.
The considered tracks are required to pass through the acceptance region of the DUT, excluding one column/row at the pixel edge as well as masked pixels and their direct neighbours.
The statistical uncertainty is calculated using a Clopper-Pearson interval of one sigma~\cite{clopper_pearson} and the systematic uncertainty arises from the threshold calibration as mentioned above. 

\paragraph{Spatial resolution}
The unbiased spatial residuals are calculated as the difference between the reconstructed cluster position and the track intercept on the DUT.
The RMS of the central $3\,\sigma$ of the distribution is extracted and the spatial telescope track resolution of \SI{2.4}{\micro \meter} for the close and \SI{5.6}{\micro \meter} for the wide telescope configuration is quadratically subtracted, which yields the spatial resolution of the DUT.

At the minimum operation threshold, the statistical uncertainty on the spatial resolution is of the order of $10^{-2}\SI{}{\micro \meter}$. 
The systematic uncertainties result from uncertainties in the telescope single-plane resolution given in~\cite{Jansen:2016bkd}.
In addition, the plane positions in z-direction are shifted independently by $\pm \SI{1}{\milli \meter}$ and the calculation of the track resolution at the position of the DUT is repeated.
Propagating the deviations to the spatial resolution yields an uncertainty of $\pm \SI{0.1}{\micro \meter}$.
The propagated threshold uncertainty evaluates to $\pm \SI{0.1}{\micro \meter}$ as well and the total systematic uncertainty is given by the quadratic sum of the two.

\paragraph{Time resolution}

Similar to the spatial residuals, the time residuals are defined as the difference between the DUT timestamp and the track timestamp. 
Signal-dependent time-walk effects are corrected by exploiting the ToT information. 
The mean time difference between the DUT and the track timestamp are subtracted for each ToT bin separately.
After correction, the RMS of the central $3\,\sigma$ of the time residuals distribution is calculated and the track time resolution of 1.1\,ns~\cite{Pitters:2019yzg} is quadratically subtracted.  

The statistical uncertainties are of the order of 0.01\,ns.
The systematic uncertainties are composed of the threshold uncertainty evaluating to $\pm 0.1$\,ns and sub-pixel by sub-pixel variations.
To quantify the latter, the analysis is repeated for every sub-pixel in a detection channel individually and the spread of the time resolution is used to define the systematic uncertainty, which yields $\pm 0.1$\,ns at the minimum operation threshold. 

\paragraph{Studies with inclined particle tracks}

The inclination angle of the DUT with respect to the beam is taken from the alignment procedure. 
The angle agrees with the nominal rotation angle set for the rotation stage apart from a constant offset. 
It was confirmed that the alignment has converged by manually modifying the plane orientation by $\pm 0.5^\circ$ and repeating the alignment. 
A deviation of less than $\pm 0.01^\circ$ is found with respect to the initial alignment.


\section{Performance for Perpendicular Particle Tracks}
\label{sec:performance}
First, measurement results for perpendicular beam incidence are presented.
Here, CLICTD sensors with different thicknesses and wafer materials are compared for the two different pixel flavours.
A comparison of the pixel flavours themselves can be found elsewhere~\cite{clictdTestbeam}.

\begin{table*}[t]
	\centering
	\caption{Mean cluster size (Cls. Size), spatial resolution (Spat. Res.) in row direction and time resolution (Time Res.) at the minimum operation threshold (Thd.) for both pixel flavours, different sensor thicknesses and wafer materials.
	C - continuous n-implant, S - segmented n-implant, Epi - epitaxial layer, Cz - Czochralski substrate.}
	\label{tab:performance_all}
	\begin{tabular}{ccccccc}
		\hline
		\toprule
		\textbf{Thickn. [\SI{}{\micro m}]} & \textbf{Material} & \textbf{Flavour} &  \textbf{Thd. [e]} &  \textbf{Cls. Size} &  \textbf{Spat. Res. (row) [\SI{}{\micro \meter}]} & \textbf{Time Res. [ns]} \\ 
		\midrule
		300 & Epi & C & $139^{+4}_{-5}$  & $1.99 \pm 0.01$ & $4.6 \pm 0.2$ & $6.5 \pm 0.1$  \\ 
		100 & Epi & C  & $136^{+4}_{-5}$  & $1.94 \pm 0.01$ & $4.6 \pm 0.2$ & $6.4 \pm 0.1$ \\ 
		50 & Epi & C  & $140^{+4}_{-5}$  & $1.91 \pm 0.01$ & $4.6 \pm 0.2$ & -   \\ 
		40 & Epi & C & $138^{+4}_{-5}$  & $1.86 \pm 0.01$ & $4.9 \pm 0.2$ & $6.4 \pm 0.1$   \\
		\midrule 
		300 & Epi & S & $136^{+4}_{-5}$ & $1.82 \pm 0.01$ & $4.6 \pm 0.2$ & $5.6 \pm 0.1$  \\
		100 & Epi & S  & $140^{+4}_{-5}$  & $1.81 \pm 0.01$  & $4.5 \pm 0.2$ & $5.5 \pm 0.1$ \\
		50 & Epi & S  & $131^{+4}_{-5}$  & $ 1.83  \pm 0.01$ & $4.6 \pm 0.2$ & - \\ 
		40 & Epi & S  & $130^{+4}_{-5}$ & $1.73 \pm 0.01$ & $4.8 \pm 0.2$ & $5.3 \pm 0.1$   \\ 
		\midrule
		100 & Cz &  S & $151^{+4}_{-5}$  & $2.36 \pm 0.01$ & $3.9 \pm 0.2$ & $4.8 \pm 0.1$   \\ 
		\bottomrule
	\end{tabular}
\end{table*}

\subsection[Cluster Size]{Cluster Size}

\begin{figure*}[bpt]
	\centering
	\begin{subfigure}[t]{0.5\textwidth}
		\centering
		\includegraphics[width=\columnwidth]{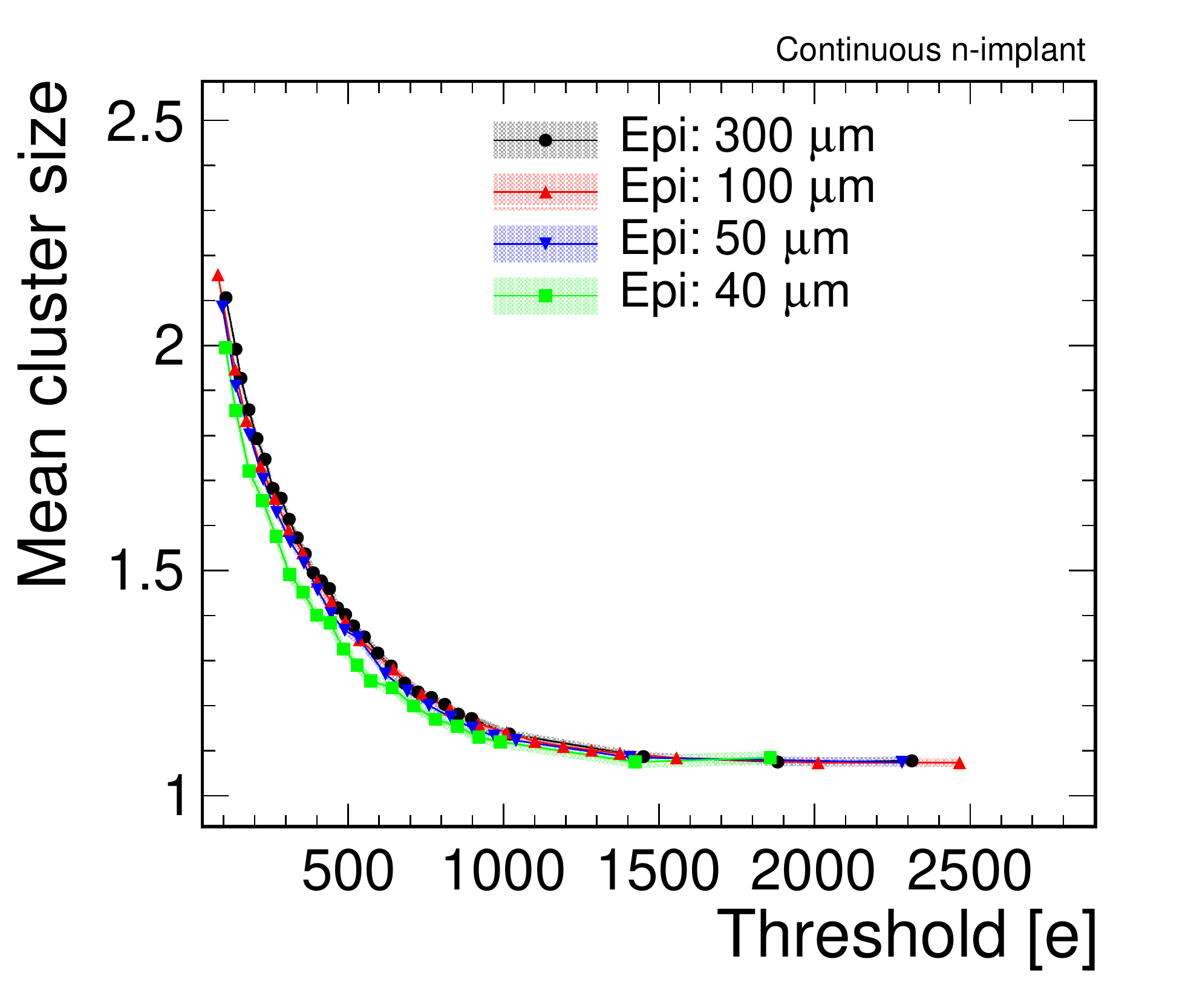}
		\caption{Continuous n-implant}
		\label{fig:meanSizeThd_modThinning}
	\end{subfigure}%
	\begin{subfigure}[t]{0.5\textwidth}
		\centering
		\includegraphics[width=\columnwidth]{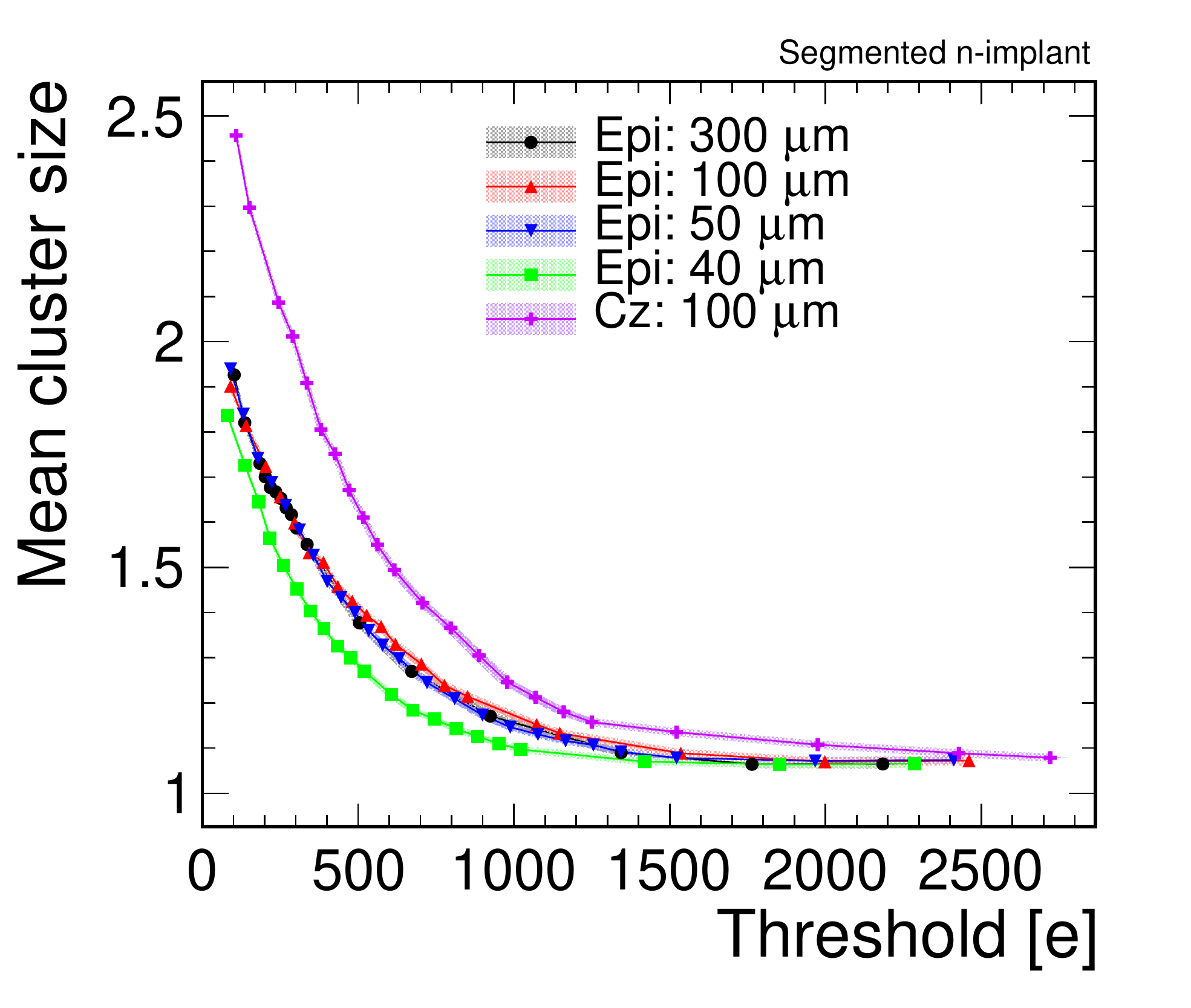}
		\caption{Segmented n-implant}
		\label{fig:meanSizeThd_gapThinning}
	\end{subfigure}%
	\caption{Mean cluster size as a function of the detection threshold using sensors with different sensor thicknesses and wafer materials for the pixel flavour with (a) continuous and (b) segmented n-implant using a bias voltage of -6\,V/-6\,V at the p-well/substrate.}
	\label{fig:meanSizeThd}
\end{figure*}

\begin{figure*}[bpt]
	\centering
	\begin{subfigure}[t]{0.5\textwidth}
		\centering
		\includegraphics[width=\columnwidth]{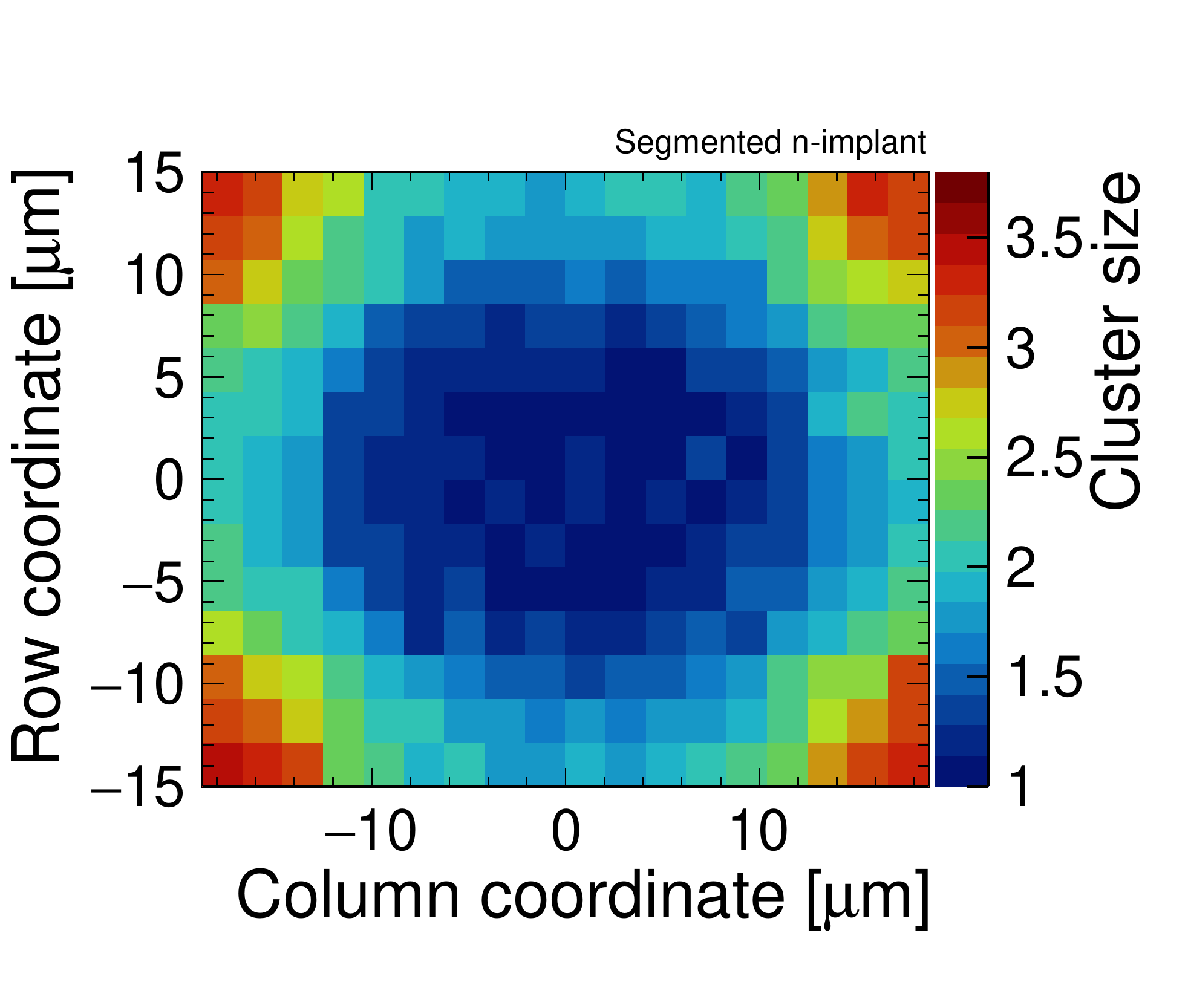}
		\caption{Epitaxial layer}
		\label{fig:clsSizeMap_epi}
	\end{subfigure}%
	\begin{subfigure}[t]{0.5\textwidth}
		\centering
		\includegraphics[width=\columnwidth]{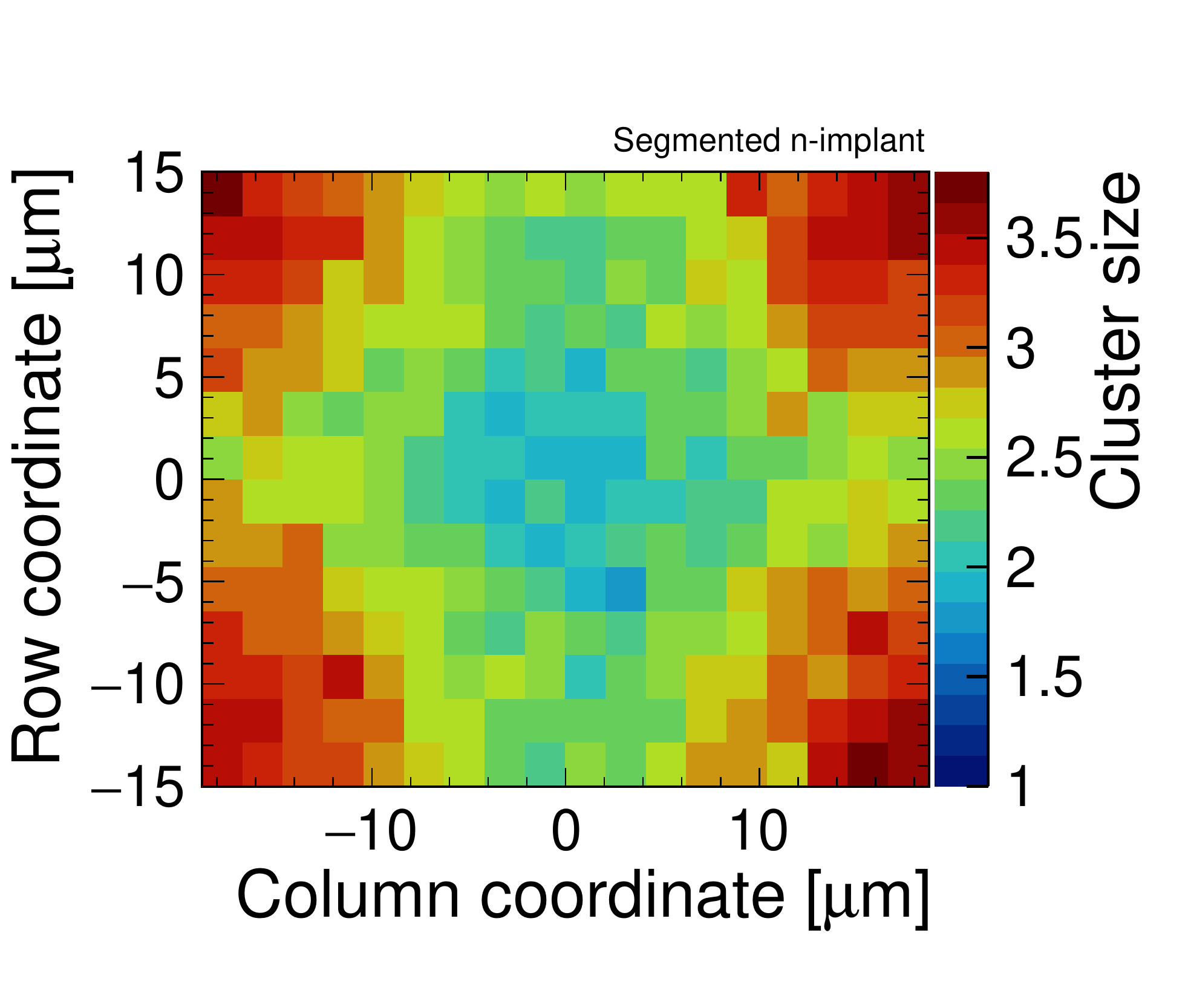}
		\caption{Czochralski substrate}
		\label{fig:clsSizeMap_cz}
	\end{subfigure}%
	\caption{In-pixel representation of the total cluster size at the minimum operation threshold for a sensor with (a) epitaxial layer and (b) Czochralski substrate.
	Both sensors have a segmented n-implant and are biased at -6\,V/-6\,V at p-wells/substrate.}
	\label{fig:sizeMap}
\end{figure*}

Comparing the cluster size of different sensors is sensitive to the total amount of induced charge and its distribution among adjacent pixel cells. 
The mean cluster size for the two pixel flavours as a function of the detection threshold is presented in Fig.~\ref{fig:meanSizeThd} and the mean size at the minimum detection threshold is listed in Table~\ref{tab:performance_all}.
The shaded band represents the uncertainties discussed in the previous section.

For both pixel flavours, the mean cluster size is the same within the uncertainties for sensor thicknesses between \SI{50}{\micro \meter} and \SI{300}{\micro \meter}.
The results imply that only the fraction of the low-resistivity substrate is removed from which charge carrier do not contribute to the measured signal. 
Thus, thinning the sensor to \SI{50}{\micro \meter} still leaves the active sensor material intact.

On the other hand, the mean cluster size for the \SI{40}{\micro \meter} thick sensor is reduced by approximately \SI{10}{\percent} at the minimum operation threshold.
As the \SI{40}{\micro \meter} thick sensor consists of approximately \SI{10}{\micro \meter} of metal layers and \SI{30}{\micro \meter} sensor material, it can be assumed that the substrate is fully removed.
Damage to the epitaxial layer by the thinning procedure~\cite{mizushima2014impact} is expected to affect the signal as well, which results in a lower cluster size. 

The decrease in mean cluster size for the \SI{40}{\micro \meter} sensors is more pronounced for the pixel flavour with segmented n-implant (cf. Fig.~\ref{fig:meanSizeThd_gapThinning}), which is consistent with the reduced charge sharing expected for this flavour.
A high degree of charge sharing leads to the distribution of the total signal to several adjacent pixel cells, thus reducing the amount of charge collected per pixel. 
In particular, charge carriers generated at the lower border of the active sensor region are subject to intense charge sharing, since their longer propagation path allows for a stronger contribution of diffusion processes. 
If the induced signal on a given pixel is not enough to surpass the threshold, the charge carriers that propagated to this cell are effectively lost. 
Therefore, this phenomenon is particularly important for the flavour with continuous n-implant and affects mostly charge carriers from the lower part of the active sensor volume. 
A removal of this volume is thus less severe, since a fraction of charge carriers are anyway lost due to sub-threshold effects. 
The stronger concentration of charge carriers for the pixel flavour with segmented n-implant mitigates the charge-sharing-induced signal loss and this flavour is consequently more sensitive to the thinning.

The mean cluster size for a \SI{100}{\micro \meter} thick sensor fabricated on a Czochralski substrate is shown in Fig.~\ref{fig:meanSizeThd_gapThinning}. 
At the minimum threshold, the mean cluster size is increased by approximately \SI{30}{\percent} compared to sensors with epitaxial layer.
The in-pixel representation of the cluster size allows for a detailed investigation of the cluster size difference, as presented in Fig~\ref{fig:sizeMap}.
In this representation, the cluster size is depicted as a function of the particle incident position within the pixel cell by folding data from a full CLICTD pixel matrix into a single cell.
The largest clusters originate from the pixel corners owing to geometrical effects and the low electric field in this region resulting in a high contribution from charge carrier diffusion.
For the sensor fabricated on Czochralski substrate, the cluster size is larger regardless of the incident position.
Especially in the pixel centre, the map exhibits mean cluster size values well above one, even though the lowest degree of charge sharing is expected from this region.
The results are thus indicative of an overall higher signal resulting from a larger active sensor volume. 

\begin{figure}[tbp]
	\centering
	\includegraphics[width=\columnwidth]{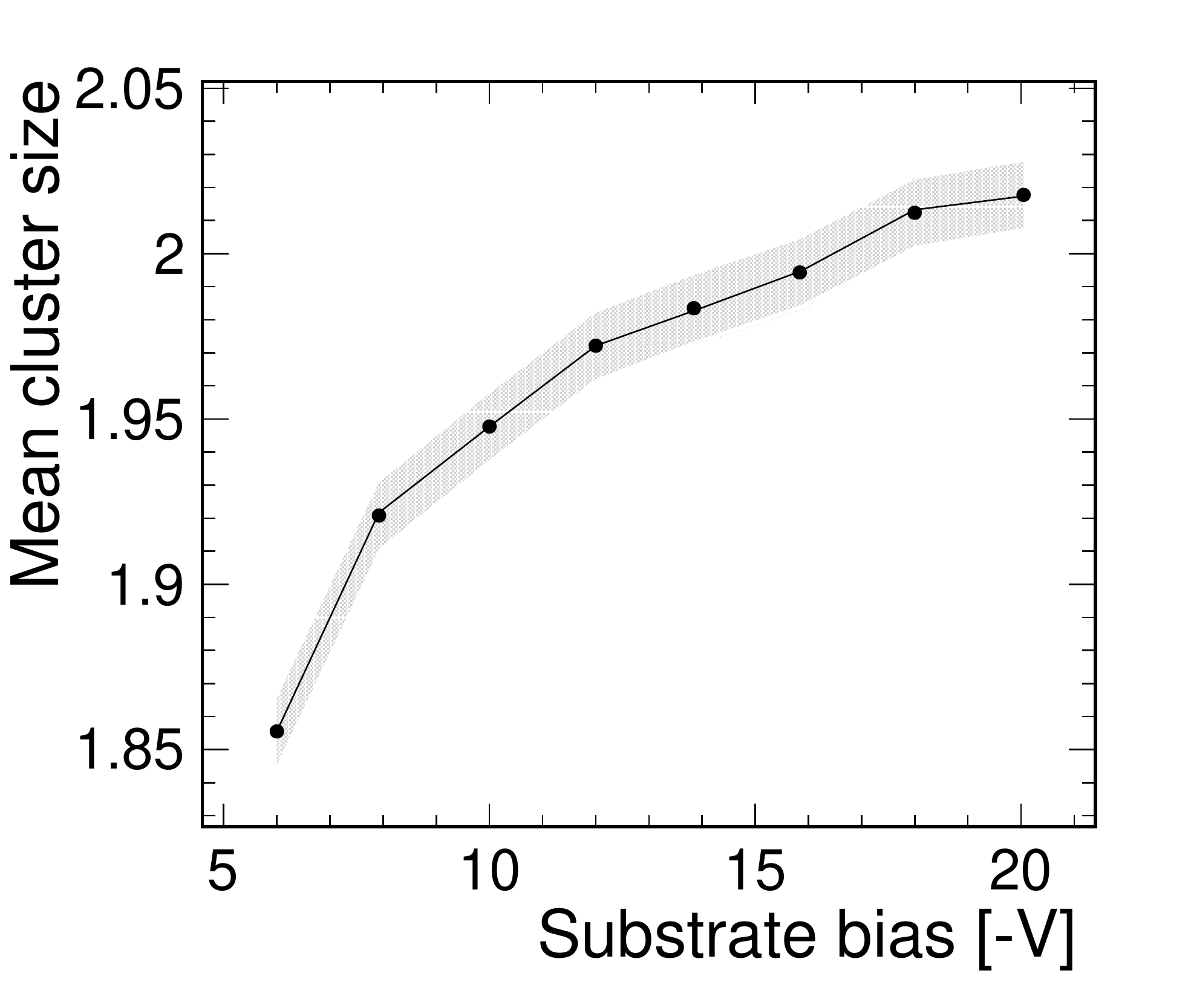}%
	\caption{Mean cluster size as a function of the substrate bias voltage at a threshold of 348\,e for a Czochralski sensor with segmented n-implant.
	The p-well voltage is fixed to -6\,V.}
	\label{fig:sizeVsBias_gap_202111}
\end{figure}

The depletion region within the Czochralski substrate is not expected to extend to the sensor backside at a bias voltage of -6\,V/-6\,V, which still limits the active sensor depth. 
An increase in substrate bias voltage, increases the depletion depth and therefore also affects the active depth, as illustrated in Fig.~\ref{fig:sizeVsBias_gap_202111}, where the mean cluster size as a function of the substrate bias voltage is displayed for the pixel flavour with segmented n-implant. 
The p-well voltage is fixed to -6 V and a higher detection threshold of 348\,e is applied to the  sensor due to the different front-end operation settings as explained before.

\subsection{Hit-Detection Efficiency}

\begin{figure*}[bpt]
	\centering
	\begin{subfigure}[t]{0.5\textwidth}
		\centering
		\includegraphics[width=\columnwidth]{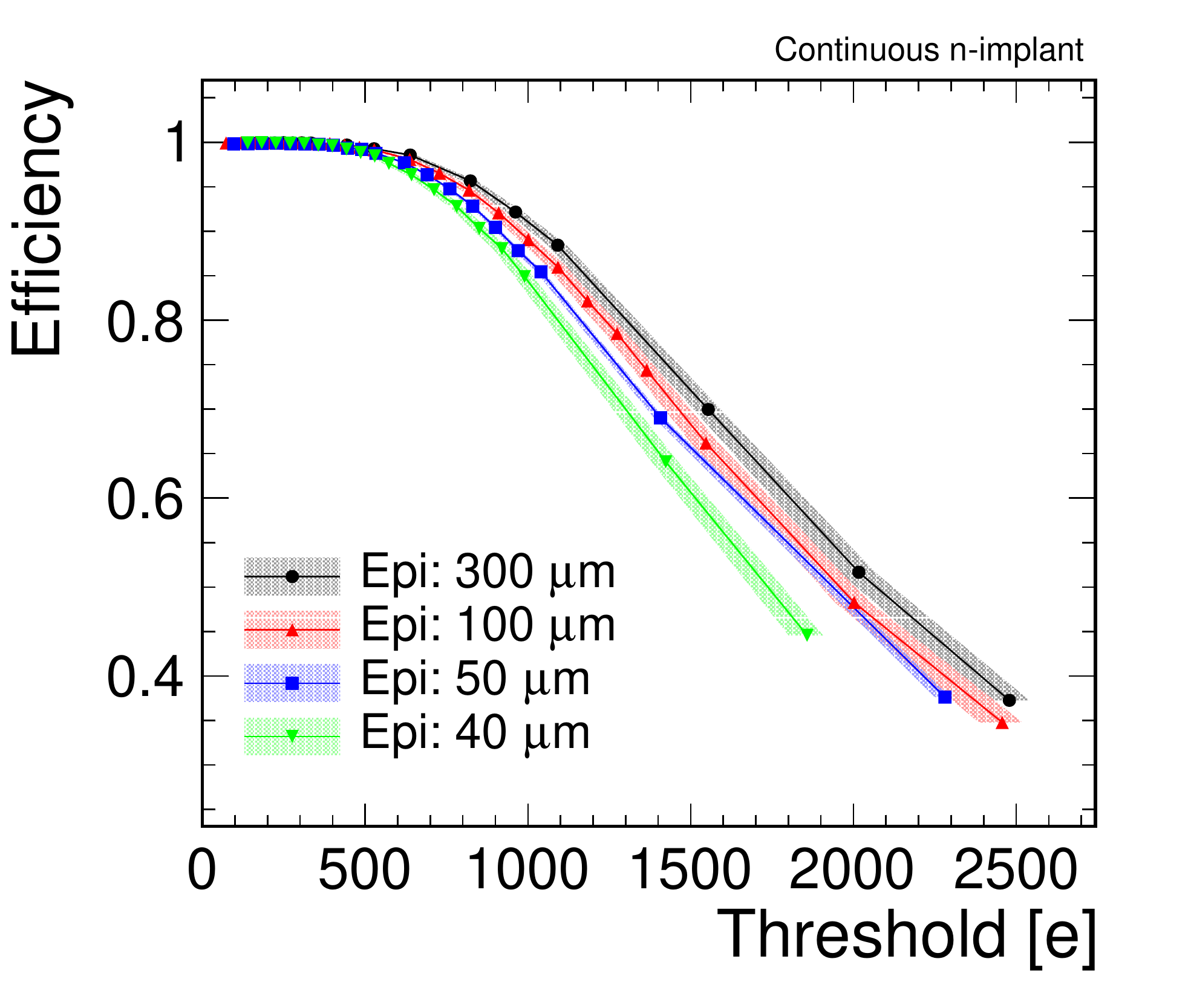}
		\caption{Continuous n-implant}
		\label{fig:effVsThd_modThinning}
	\end{subfigure}%
	\begin{subfigure}[t]{0.5\textwidth}
		\centering
		\includegraphics[width=\columnwidth]{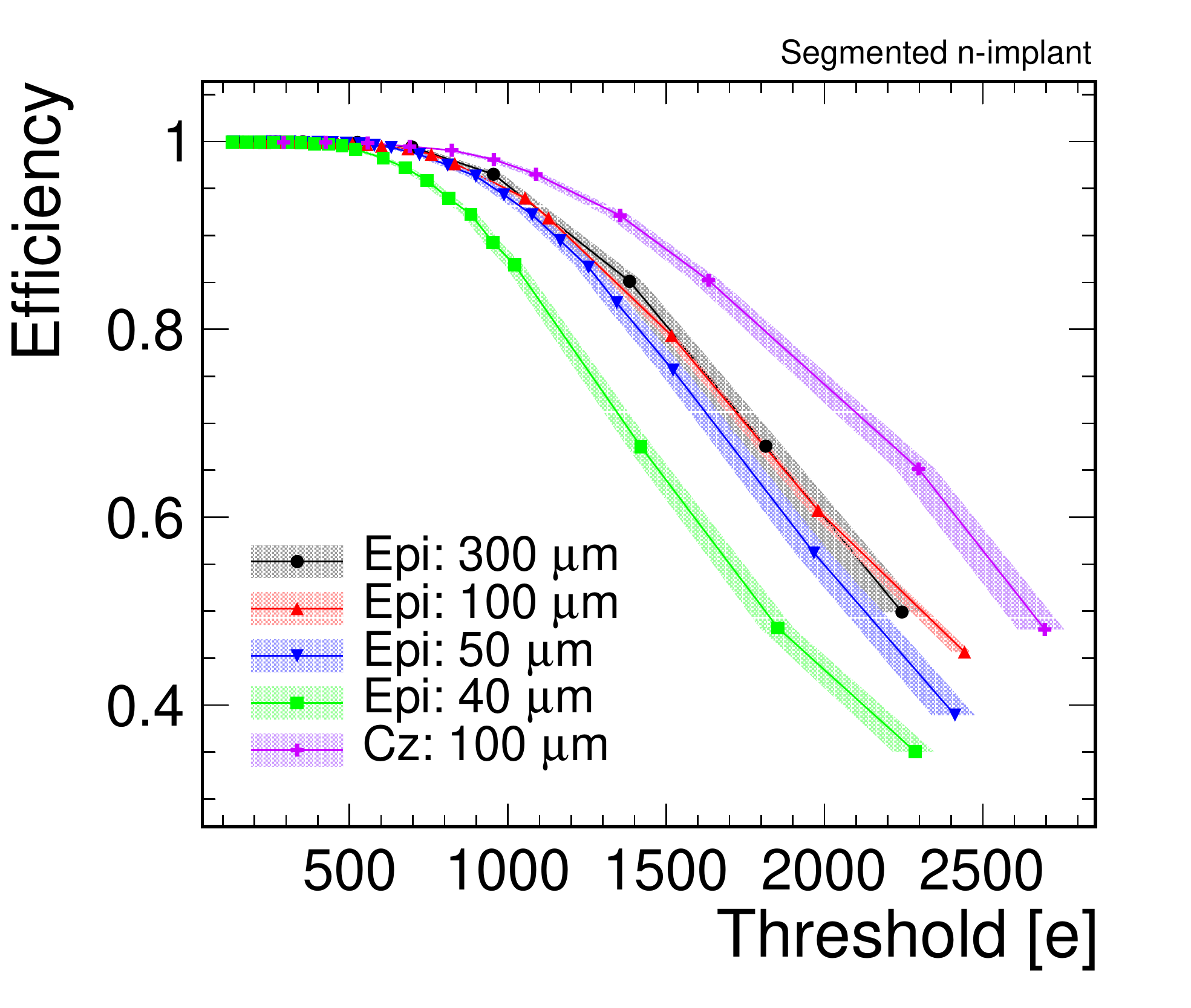}
		\caption{Segmented n-implant}
		\label{fig:effVsThd_gapThinning}
	\end{subfigure}%
	\caption{Hit-detection efficiency as a function of the detection threshold using sensors with different sensor thicknesses and wafer materials for the pixel flavour with (a) continuous and (b) segmented n-implant using a bias voltage of -6\,V/-6\,V at the p-well/substrate.}
	\label{fig:effVsThd}
\end{figure*}

The hit-detection efficiency is closely related to the maximum single-pixel charge (\textit{seed charge}) in a cluster and is thus correlated with the total signal and the degree of charge sharing. 
The efficiency is determined as a function of the detection threshold as presented in Fig.~\ref{fig:effVsThd} for both pixel flavours. 
While efficiencies well above \SI{99}{\percent} are achieved at low detection thresholds, the efficiency deteriorates for values greater than 500\,e, since all single-pixel signals in a cluster can fall below the detection threshold.
The degradation is stronger for the pixel flavour with continuous n-implant due to the enhanced charge sharing, which leads to a smaller charge per pixel, as discussed in detail in~\cite{clictdTestbeam}.

\begin{figure*}[bpt]
	\centering
	\begin{subfigure}[t]{0.5\textwidth}
		\centering
		\includegraphics[width=\columnwidth]{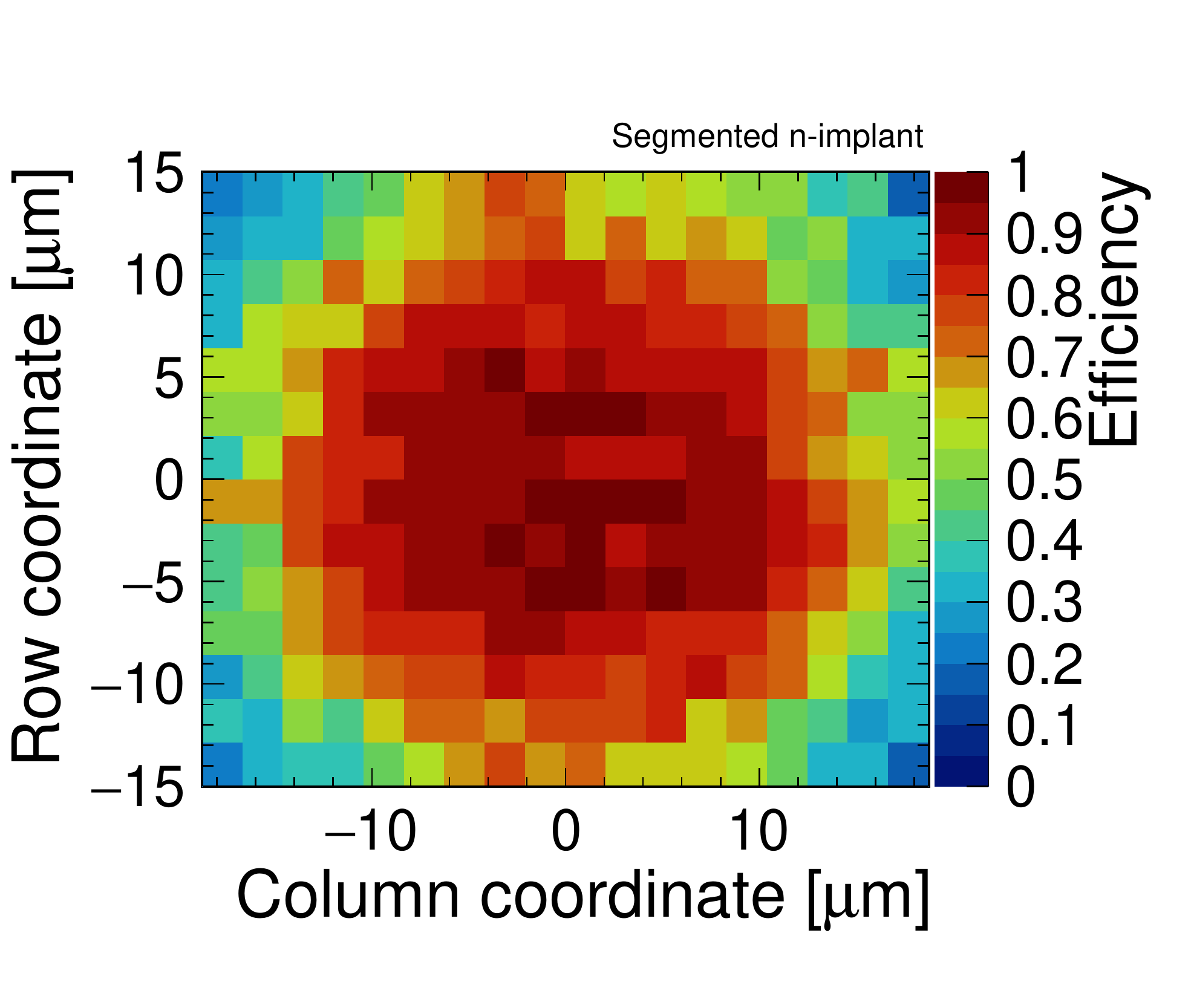}
		\caption{Epitaxial layer}
		\label{fig:effMap_contN}
	\end{subfigure}%
	\begin{subfigure}[t]{0.5\textwidth}
		\centering
		\includegraphics[width=\columnwidth]{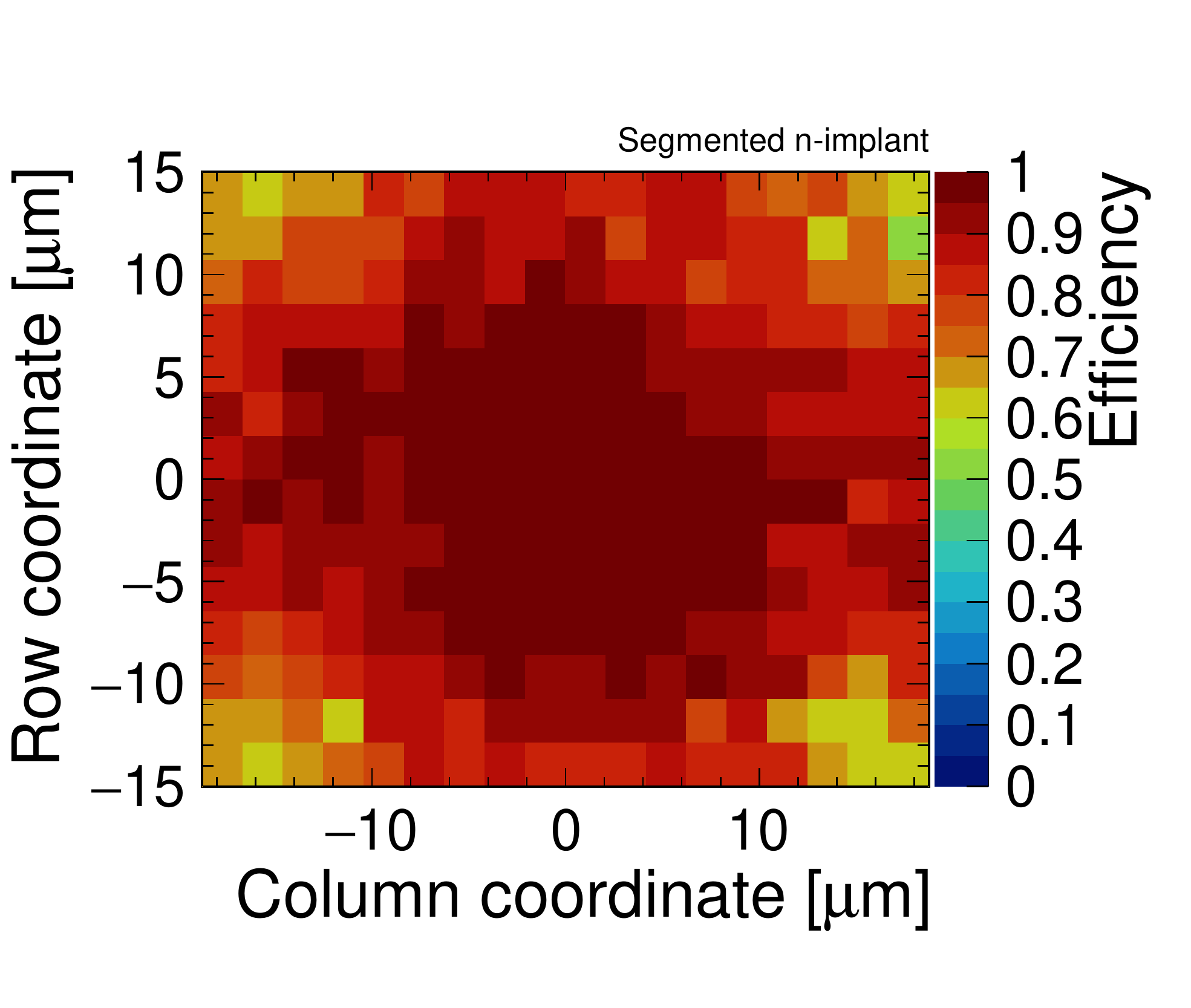}
		\caption{Czochralski substrate}
		\label{fig:effMap_cz}
	\end{subfigure}%
	\caption{In-pixel representation of the hit-detection efficiency at a threshold of 1950\,e for a sensor with (a) epitaxial layer and (b) Czochralski substrate.
		Both sensors have a segmented n-implant and are biased at -6\,V/-6\,V at p-wells/substrate.}
	\label{fig:effMap}
\end{figure*}

For high thresholds, inefficient regions start to form at the pixel borders, as illustrated in Fig.~\ref{fig:effMap_contN}, where the in-pixel efficiency is shown at a threshold of 1950\,e for a \SI{300}{\micro \meter} thick sensor with segmented n-implant and epitaxial layer.
As the diffusion of charge carriers to neighbouring pixels is enhanced at the edges, a smaller seed signal and consequently a lower efficiency is associated with these regions.

For the \SI{40}{\micro m} thick sensors, the high-efficiency plateau is noticeably reduced compared to the thicker sensors.
In agreement with the smaller cluster size observed in the previous section, the degraded efficiency indicates an overall reduction in signal compared to the thicker sensors.
These results support the assumption of a smaller active depth due to the removal of sensitive sensor volume.
The degradation in efficiency is less severe for the pixel flavour with continuous n-implant as discussed above.  
A slight trend towards smaller efficiencies is also visible for \SI{50}{\micro m} thick sensors, although it is covered by the systematic uncertainties. 
The results indicate that parts of the active material are potentially already damaged in the \SI{50}{\micro m} thick sensors.

The sensor fabricated on a Czochralski substrate exhibits a larger efficiency at high detection thresholds compared to sensors with epitaxial layer as a direct consequence of the higher signal. 
The in-pixel representation of the efficiency is depicted in Fig.~\ref{fig:effMap_cz} at a detection threshold of approximately 1950\,e and confirms that the efficiency is larger especially in the pixel edges, where the highest degree of charge sharing is expected.

\begin{figure}[tbp]
	\centering
	\includegraphics[width=\columnwidth]{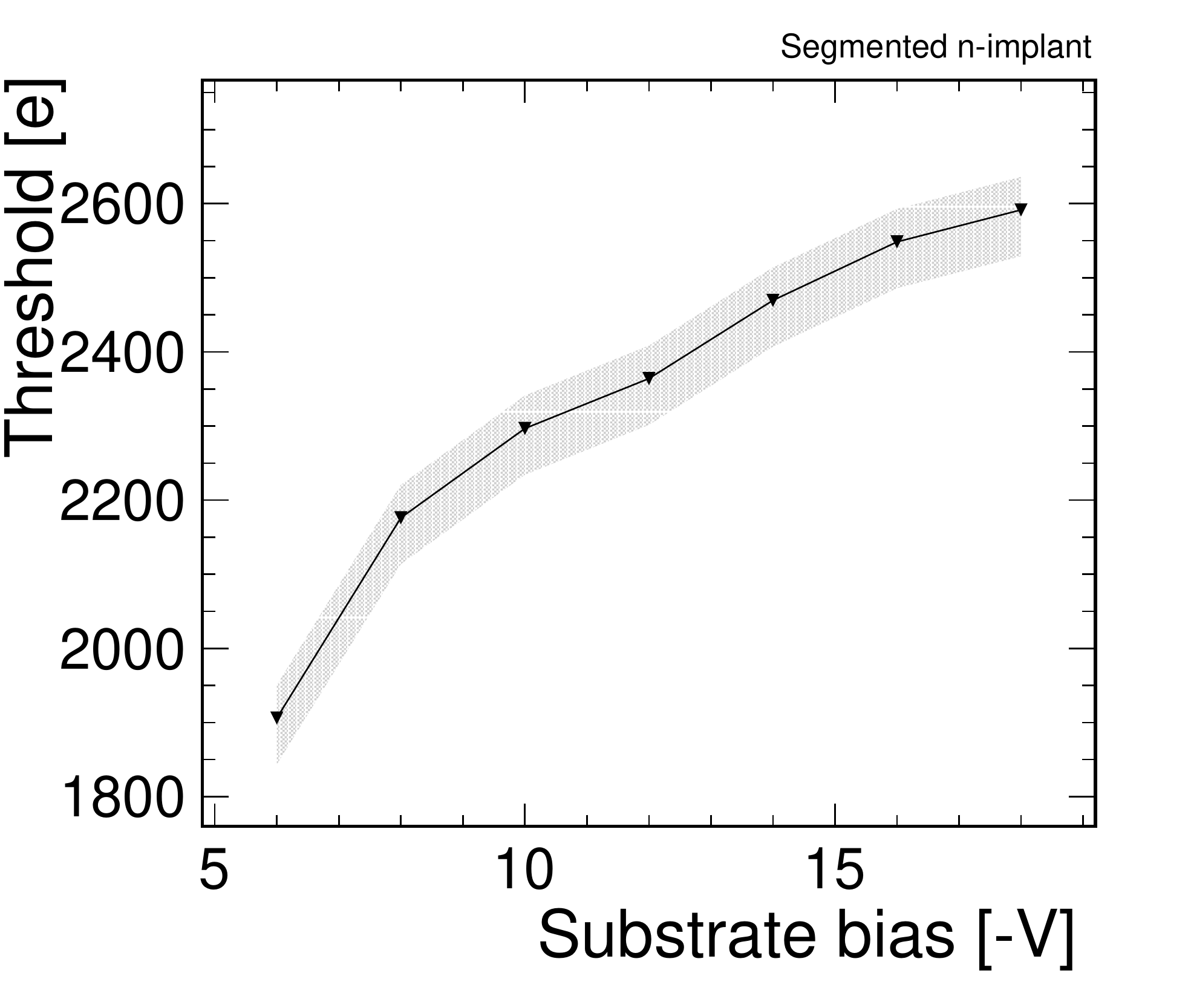}%
	\caption{Detection threshold corresponding to an efficiency of \SI{80}{\percent} as a function of the substrate bias voltage for a Czochralski sample with segmented n-implant.
		The p-well voltage is fixed to -6\,V.}
	\label{fig:effVsSubstrate_gapN}
\end{figure}

The impact of the substrate voltage for samples with Czochralski substrate is illustrated in Fig.~\ref{fig:effVsSubstrate_gapN}, where the detection threshold corresponding to an efficiency of \SI{80}{\percent} is presented as a function of the substrate bias voltage.
The threshold increases by about \SI{30}{\percent} from -6\,V to -20\,V. 
At -20\,V, the value is about \SI{85}{\percent} higher compared to the corresponding threshold for samples with epitaxial layer, which evaluates to approximately 1400\,e.

\subsection[Spatial Resolution]{Spatial Resolution}

\begin{figure*}[bpt]
	\centering
	\begin{subfigure}[t]{0.5\textwidth}
		\centering
		\includegraphics[width=\columnwidth]{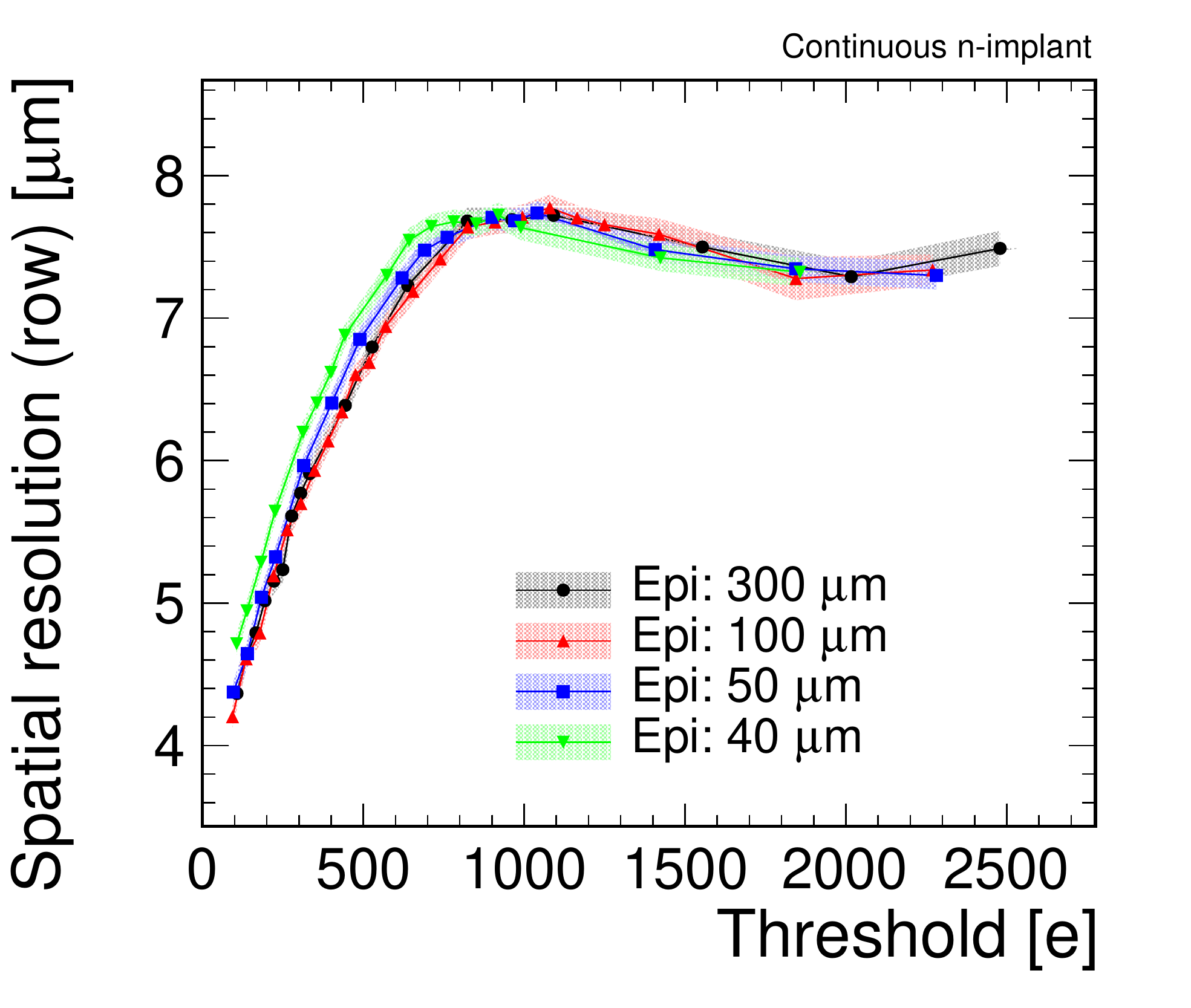}
		\caption{Continuous n-implant}
		\label{fig:sRThd_contThinning}
	\end{subfigure}%
	\begin{subfigure}[t]{0.5\textwidth}
		\centering
		\includegraphics[width=\columnwidth]{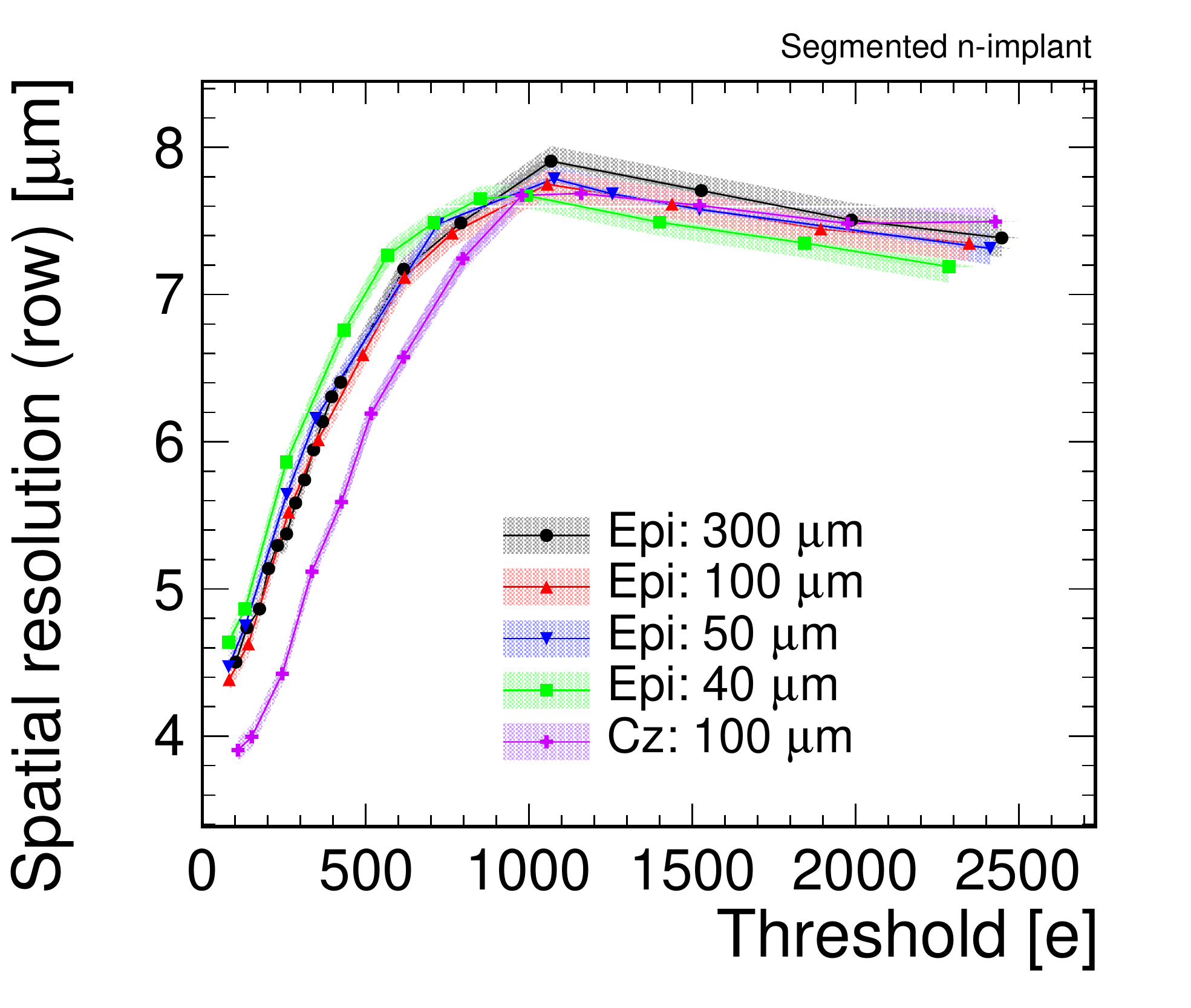}
		\caption{Segmented n-implant}
		\label{fig:effVsThd_gapThinning}
	\end{subfigure}%
	\caption{Spatial resolution as a function of the detection threshold using sensors with different thicknesses and wafer materials for the pixel flavour with (a) continuous and (b) segmented n-implant using a bias voltage of -6\,V/-6\,V at the p-well/substrate.}
	\label{fig:sRThd}
\end{figure*}

\begin{figure}[tbp]
	\centering
	\includegraphics[width=\columnwidth]{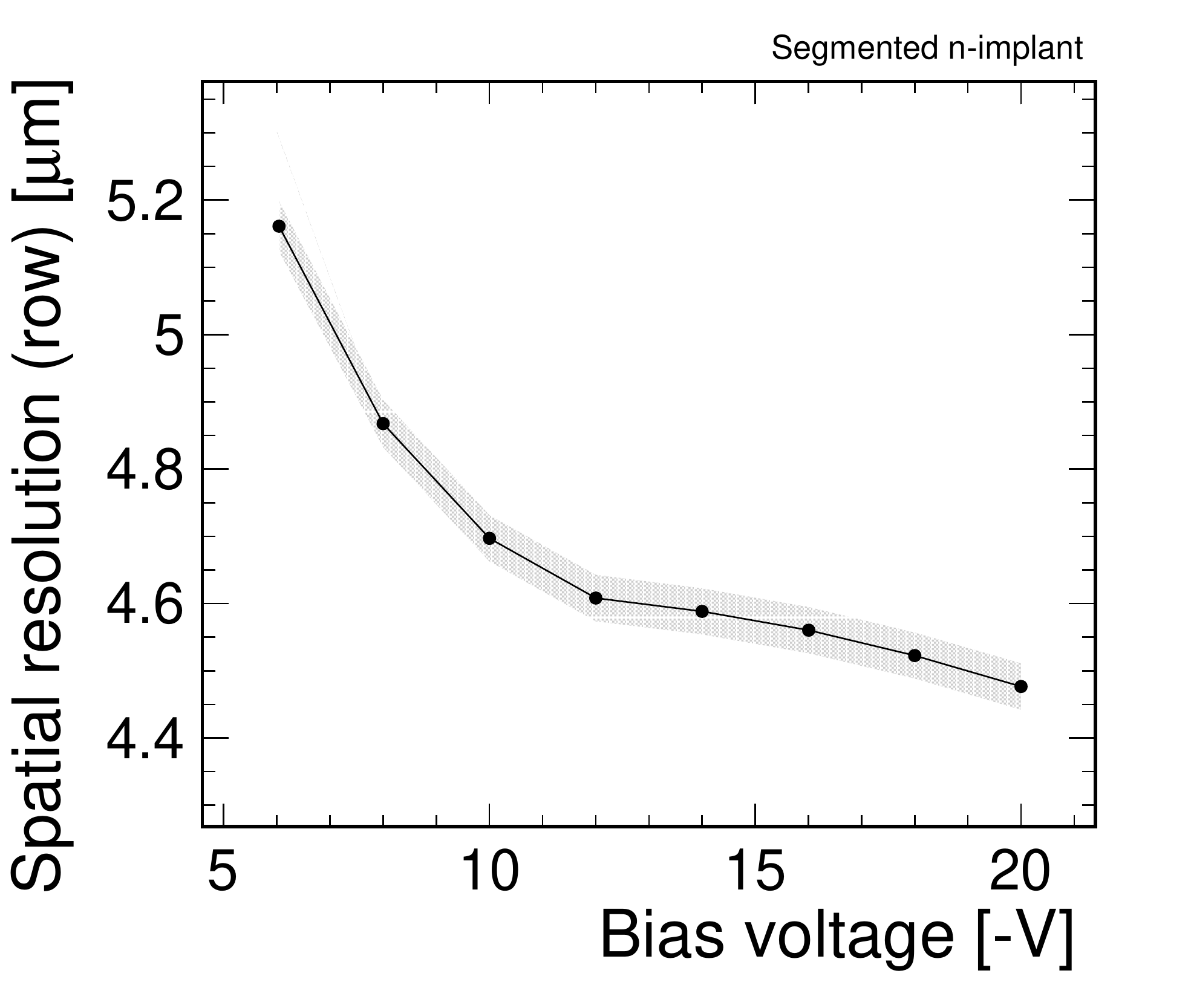}%
	\caption{Spatial resolution as a function of the substrate bias voltage at a threshold of 348\,e for a Czochralski sample with segmented n-implant.
		The p-well voltage is fixed to -6\,V.}
	\label{fig:sRThd_CzBias_gapN}
\end{figure}

The spatial resolution in row direction as a function of the detection threshold is presented in Fig.~\ref{fig:sRThd} for both pixel flavours and the results at the minimum threshold are listed in Table~\ref{tab:performance_all}.
For thresholds above 1200\,e, no $\eta$-correction is applied, since the application of the algorithm becomes challenging due to the small number of two-pixel clusters. 

As the modifications to the n-implant are not applied in row direction, the charge sharing behaviour is similar for both pixel flavours and the spatial resolution is thus in good agreement within the uncertainties.
Although the resolution degrades with increasing threshold due to the decrease in cluster size, the binary resolution of \SI{8.7}{\micro \meter} is never exceeded. 
For high threshold values, an improvement of the spatial resolution is caused by the formation of inefficient regions at the pixel edges, as displayed in Fig.~\ref{fig:effMap_contN}. 
These inefficiencies lead to an effectively smaller pixel pitch that results in an artificial improvement in spatial resolution.

Within the uncertainties, the spatial resolution for the $\geq \SI{50}{\micro \meter}$ thick sensors are in good agreement owing to the similar cluster size at a given threshold.

The spatial resolution of the \SI{40}{\micro \meter} thick sensor degrades for thresholds smaller than $1000$\,e owing to the smaller cluster size at a given threshold (cf. Fig.~\ref{fig:meanSizeThd}).
For the flavour with continuous n-implant, the degradation is as high as \SI{7}{\percent} at the minimum detection threshold.
The difference vanishes at high thresholds, where single-pixel clusters dominate for all sensor thicknesses.

The higher signal from the Czochralski sensors leads to a larger cluster size and consequently an improved spatial resolution. 
The difference is particularly noticeable at small threshold values in accordance with the larger difference in cluster size that was presented in Fig.~\ref{fig:meanSizeThd_gapThinning}. 
At the minimum operation threshold listed in Table~\ref{tab:performance_all}, the resolution improves by about \SI{15}{\percent}. 
At high thresholds, the mean cluster size converges to one resulting in an identical resolution within the uncertainties.

With increasing substrate bias voltage, the depleted region expands evoking a higher signal that leads to a larger cluster size and consequently an improved spatial resolution, as illustrated in Fig.~\ref{fig:sRThd_CzBias_gapN} for a Czochralski sensor with segmented n-implant at a comparably high threshold of 348\,e. 
Between -6\,V and -20\,V, the spatial resolution improves by approximately \SI{13}{\percent}.
While the comparably high threshold limits the absolute performance improvement, the potential of the Czochralski substrate is still distinguishable.

\subsection{Time Resolution}

\begin{figure*}[bpt]
	\centering
	\begin{subfigure}[t]{0.5\textwidth}
		\centering
		\includegraphics[width=\columnwidth]{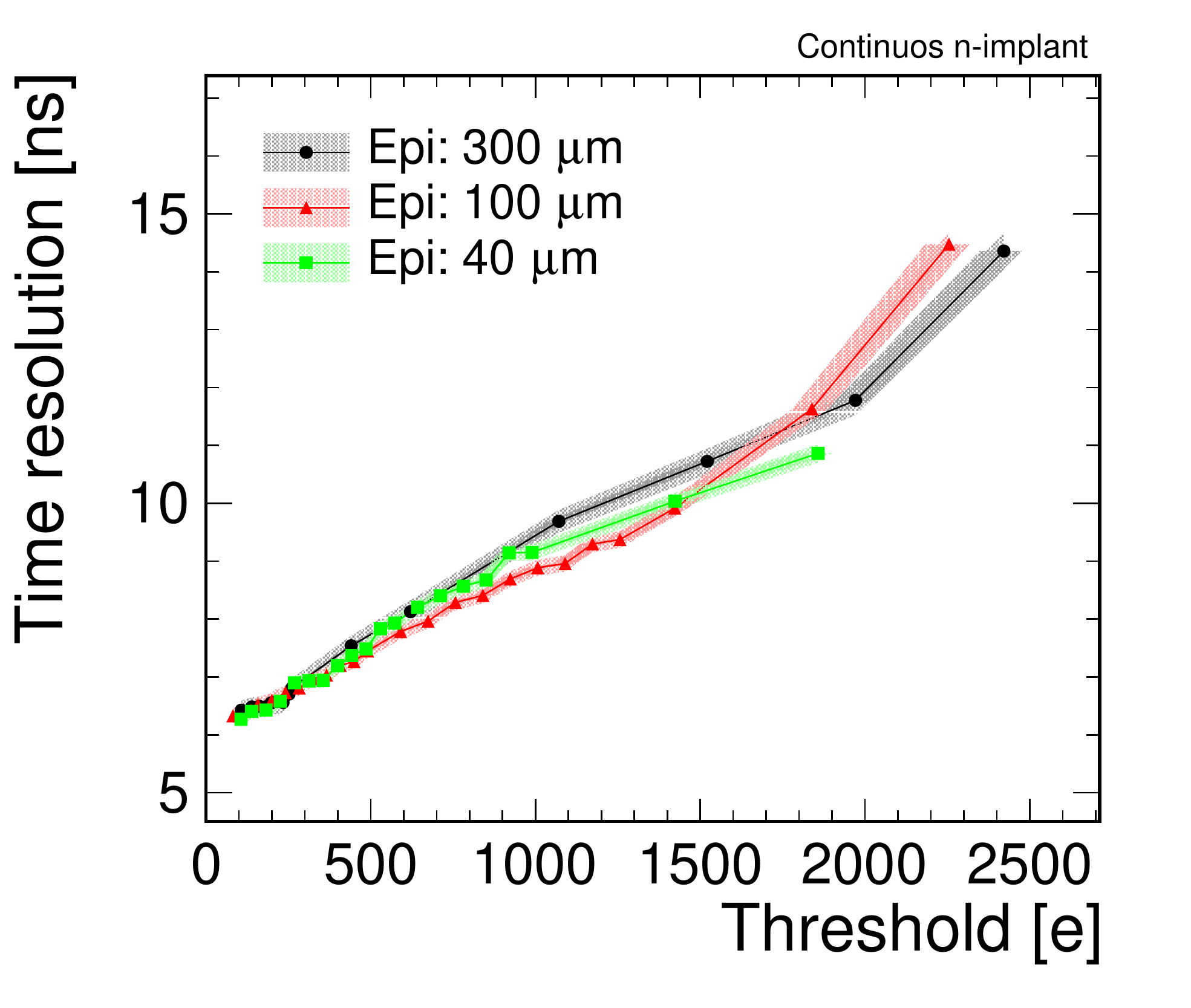}
		\caption{Continuous n-implant}
		\label{fig:tS_modThinning_202112}
	\end{subfigure}%
	\begin{subfigure}[t]{0.5\textwidth}
		\centering
		\includegraphics[width=\columnwidth]{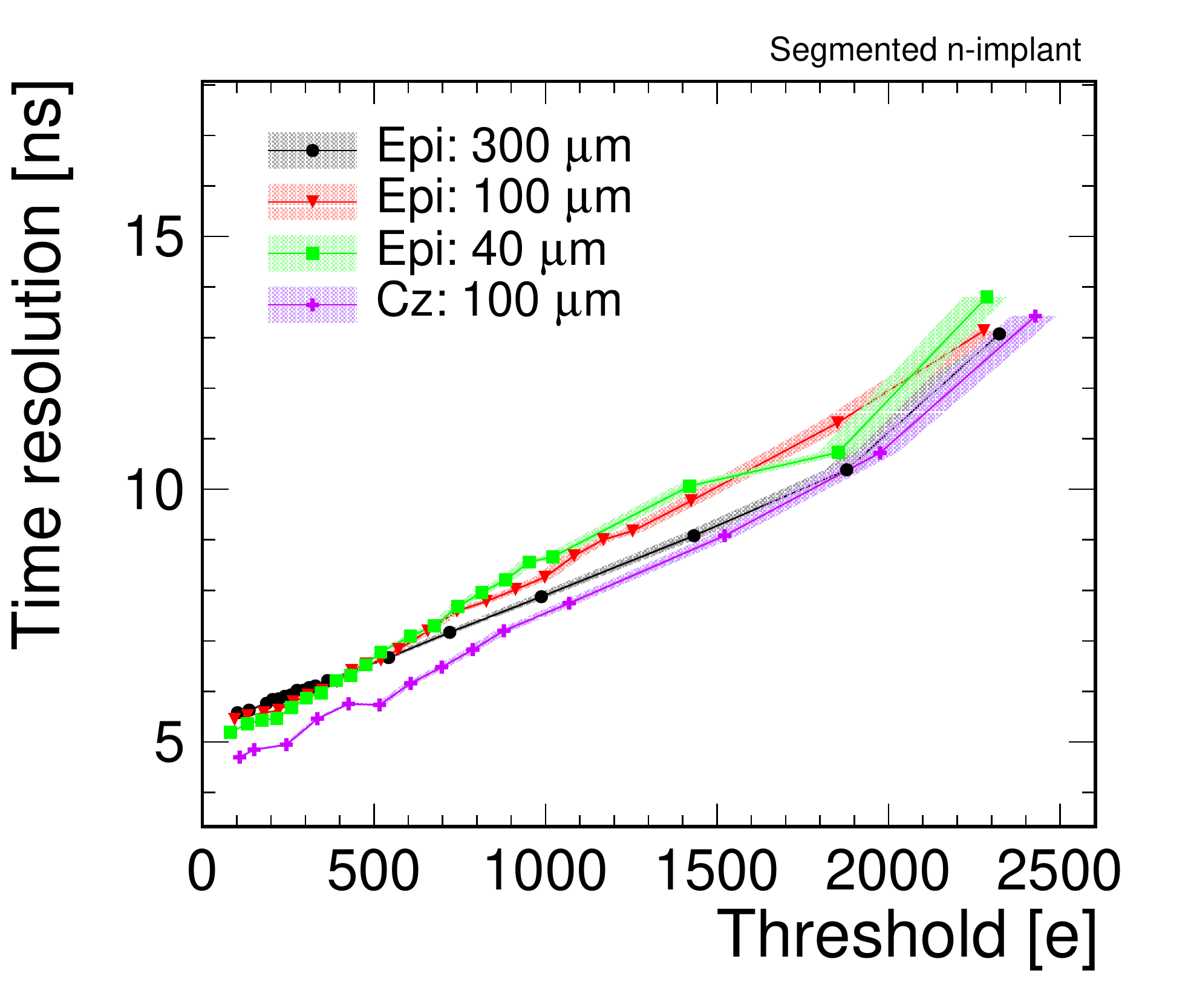}
		\caption{Segmented n-implant}
		\label{fig:timeVsThd_thinnedWCz}
	\end{subfigure}%
	\caption{Time resolution as a function of the detection threshold using sensors with different thicknesses and wafer materials for the pixel flavour with (a) continuous and (b) segmented n-implant using a bias voltage of -6\,V/-6\,V at the p-well/substrate.}
	\label{fig:imeVsThd}
\end{figure*}

The time resolution after time-walk correction is depicted in Fig.~\ref{fig:imeVsThd} as a function of the detection threshold for both pixel flavours. 
The results at the minimum operation threshold are listed in Table~\ref{tab:performance_all}. 
With increasing threshold, the time resolution degrades owing to a stronger contribution of amplitude noise causing a time jitter. 
The jitter is inversely proportional to the slope of the signal at the threshold-crossing point, which flattens towards the peak of the signal. 

\begin{figure}[tbp]
	\centering
	\includegraphics[width=\columnwidth]{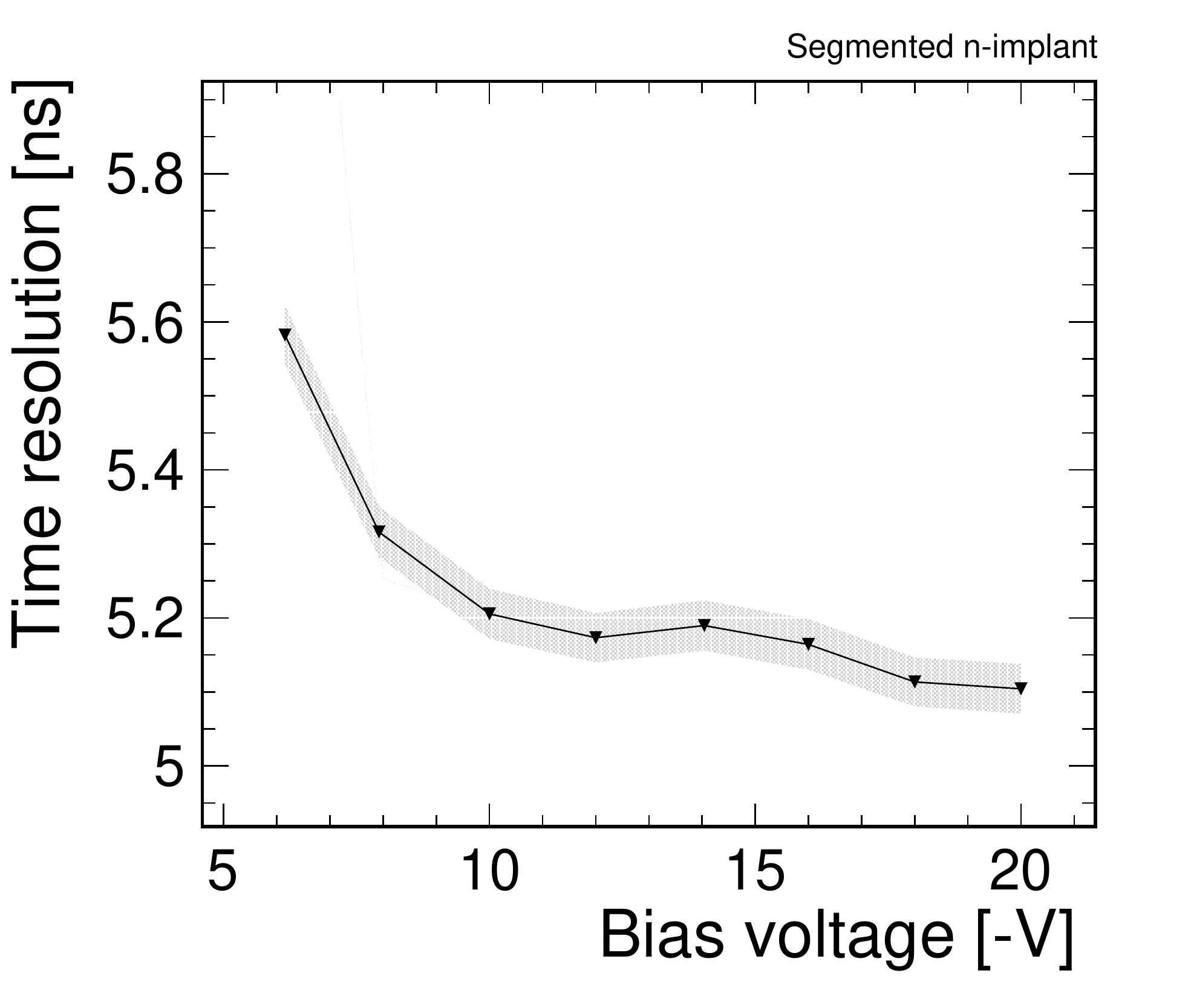}%
	\caption{Time resolution as a function of the substrate bias voltage at a threshold of 348\,e for a Czochralski sensor with segmented n-implant.
	The p-well voltage is fixed to -6\,V.}
	\label{fig:tS_CzBias_gapN}
\end{figure}

It is expected that the time resolution is limited by the binning of the ToA clock as well as the time-walk procedure~\cite{clictdTestbeam}, which obscure sensor effects related to the device thickness.
Nevertheless, a \SI{14}{\percent} improvement is visible for the Czochralski sensor owing to a larger seed signal, which facilitates the time-walk correction and suppresses time jitter.
The different front-end settings for the Czochralski sensors only concern the input node and are therefore not expected to have a significant impact on the time resolution. 
An increase in substrate bias voltage leads to an additional improvement in time resolution, as presented in Fig.~\ref{fig:tS_CzBias_gapN} at a threshold of 348\,e.
Between -6\,V and -20\,V, the time resolution improved by approximately  \SI{9}{\percent}.

\section{Studies with Inclined Particle Tracks}

In the following, the sensor performance is assessed for inclined particle tracks and the active sensor depth is investigated.

\subsection{Performance}
In many HEP applications, particles enter the sensor under an oblique angle, due to e.g. mechanical rotation of detector modules or helical particle trajectories in a magnetic field. 
Therefore, the sensor performance for inclined particle tracks merits detailed investigation.
Here, a \SI{300}{\micro \meter} thick sensor with epitaxial layer and continuous n-implant is used to exemplify the effects of the inclination angle on the sensor performance.

\paragraph{Cluster Size}

\begin{figure*}[bpt]
	\centering
	\begin{subfigure}[t]{0.5\textwidth}
		\centering
		\includegraphics[width=\columnwidth]{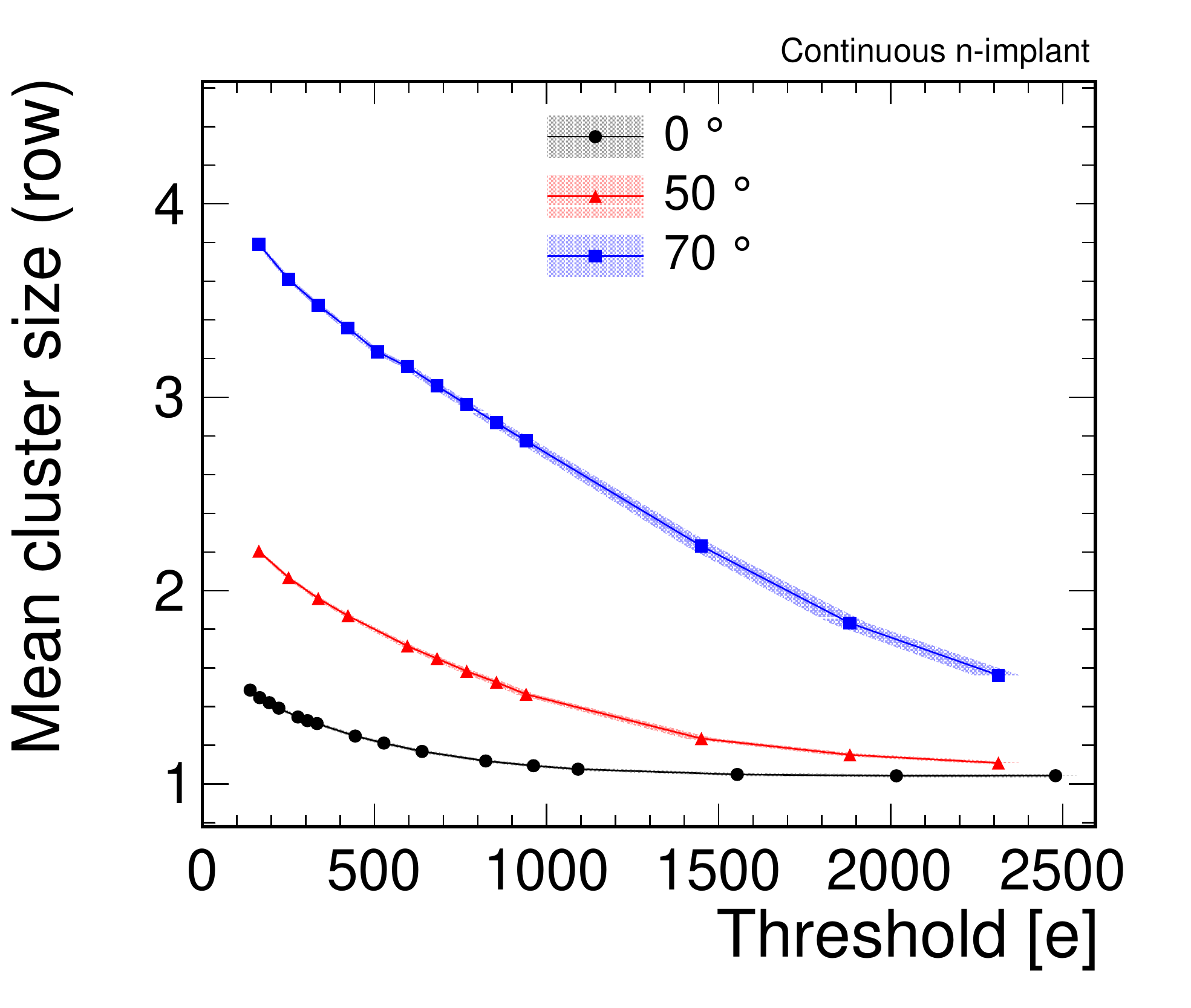}
		\caption{Cluster size in row direction}
		\label{fig:meanSizeRowThd_rotation}
	\end{subfigure}%
	\begin{subfigure}[t]{0.5\textwidth}
		\centering
		\includegraphics[width=\columnwidth]{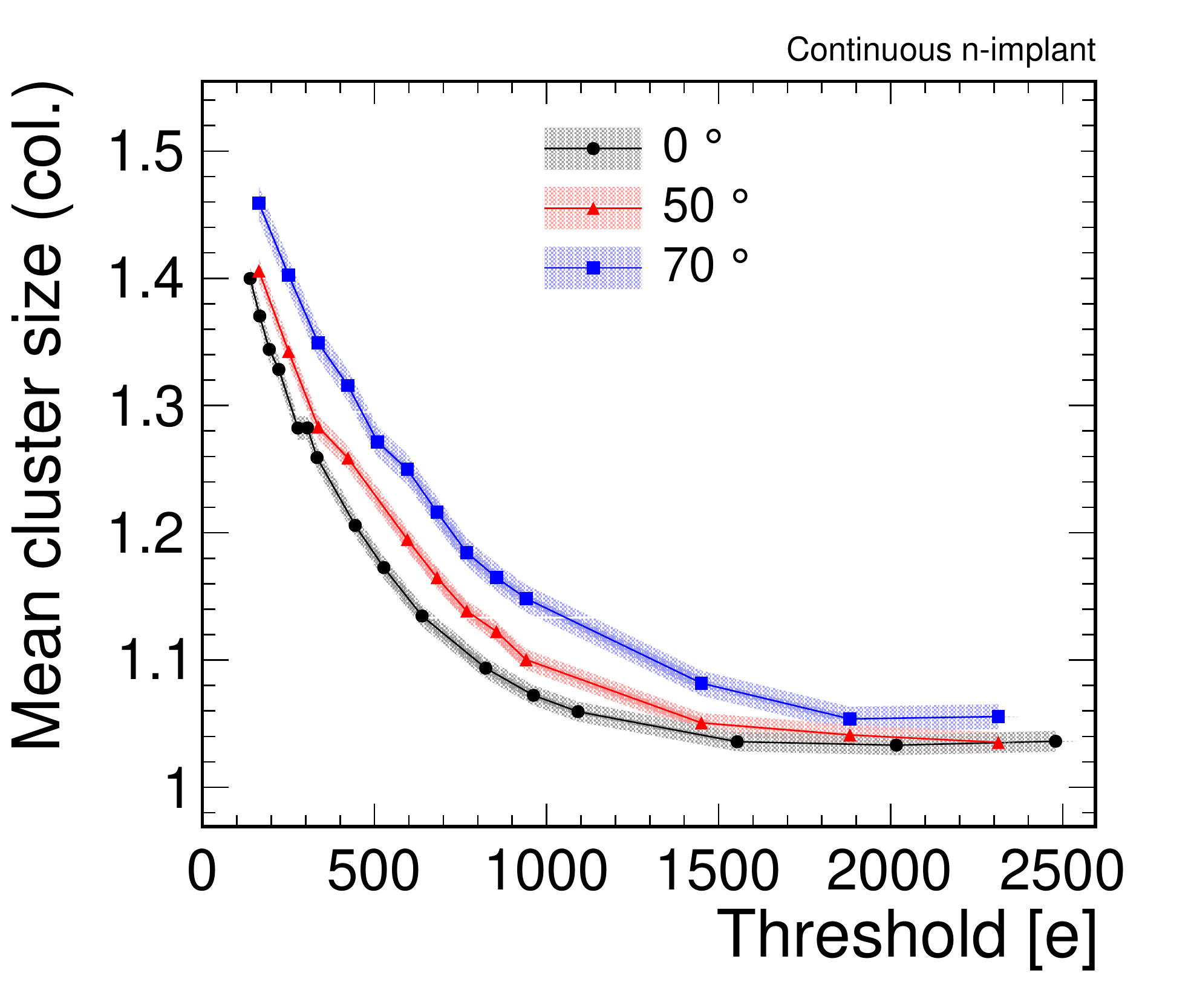}
		\caption{Cluster size in column direction}
		\label{fig:meanSizeColThd_rotation}
	\end{subfigure}%
	\caption{Cluster size as a function of the detection threshold for different rotation angles for a \SI{300}{\micro \meter} thick sensor with epitaxial layer and continuous n-implant tilted in row direction.
	A bias voltage of -6\,V/-6\,V is applied to the p-well/substrate.}
	\label{fig:meanSizeThd_rotation}
\end{figure*}

The amount of active silicon traversed by particles is varied by inclining the sensor relative to the beam axis. 
For high inclination angles, particle tracks cross several adjacent pixel cells, giving rise to a larger cluster size as illustrated in Fig.~\ref{fig:meanSizeThd_rotation} for a sensor tilted in row direction. 
The mean cluster size at the minimum detection threshold is listed in Table~\ref{tab:inclined_sensor_performance}.
A considerable increase in cluster size in row direction is distinguishable principally due to the  geometrical effect of charge deposition in several pixel cells.
Between $0^\circ$ and $70^\circ$, the increase is as high as \SI{250}{\percent} at the minimum operation threshold.
The simultaneous increase in cluster size in column direction is consistent with an overall increase in the number of liberated charge carriers, whose undirected diffusion also affects charge sharing in column direction. 
At the minimum operation threshold, the mean cluster size in column direction is approximately \SI{6}{\percent} larger at $70^\circ$ compared to perpendicular incidence.

\paragraph{Efficiency}

\begin{figure}[tbp]
	\centering
	\includegraphics[width=\columnwidth]{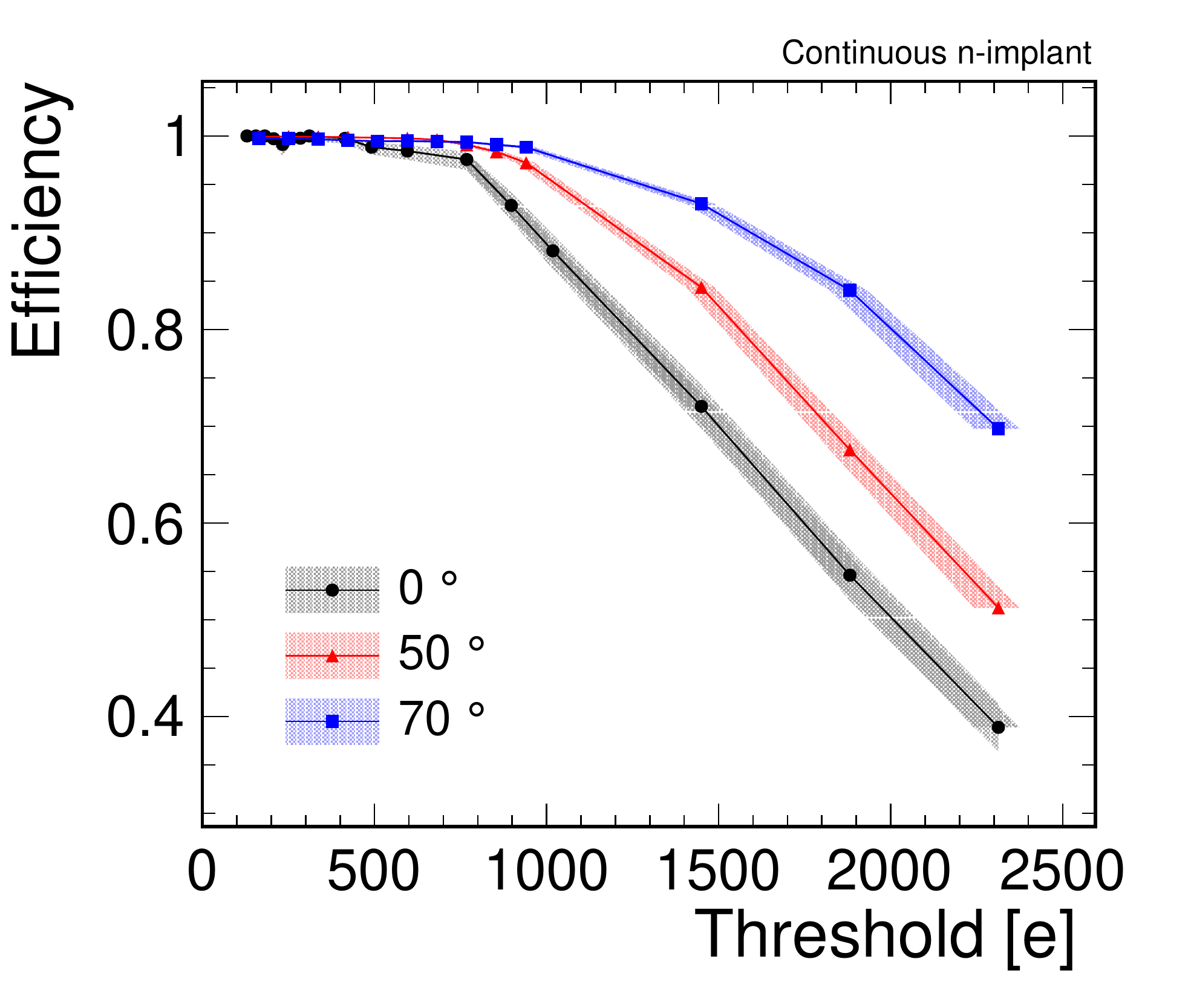}%
	\caption{Detection efficiency as a function of the detection threshold for different rotation angles for a \SI{300}{\micro \meter} thick sensor with epitaxial layer and continuous n-implant.
	The sensor was tilted in row direction and the p-well/substrate was biased at -6\,V/-6\,V.}
	\label{fig:effVsThd_rot_202110}
\end{figure}

With increasing inclination angle, the total energy deposition in the sensor increases due to the longer particle path in the active sensor region. 
As a result, a higher signal is detected, which leads to an appreciable increase in efficiency at high thresholds, as depicted in Fig.~\ref{fig:effVsThd_rot_202110}, where the efficiency as a function of the detection threshold is shown for three different rotation angles. 
At a threshold of 2300\,e, the efficiency increases from about \SI{38}{\percent} at 0$^\circ$  to \SI{70}{\percent} at  70$^\circ$.

\paragraph{Spatial Resolution}

\begin{figure}[tbp]
	\centering
	\includegraphics[width=\columnwidth]{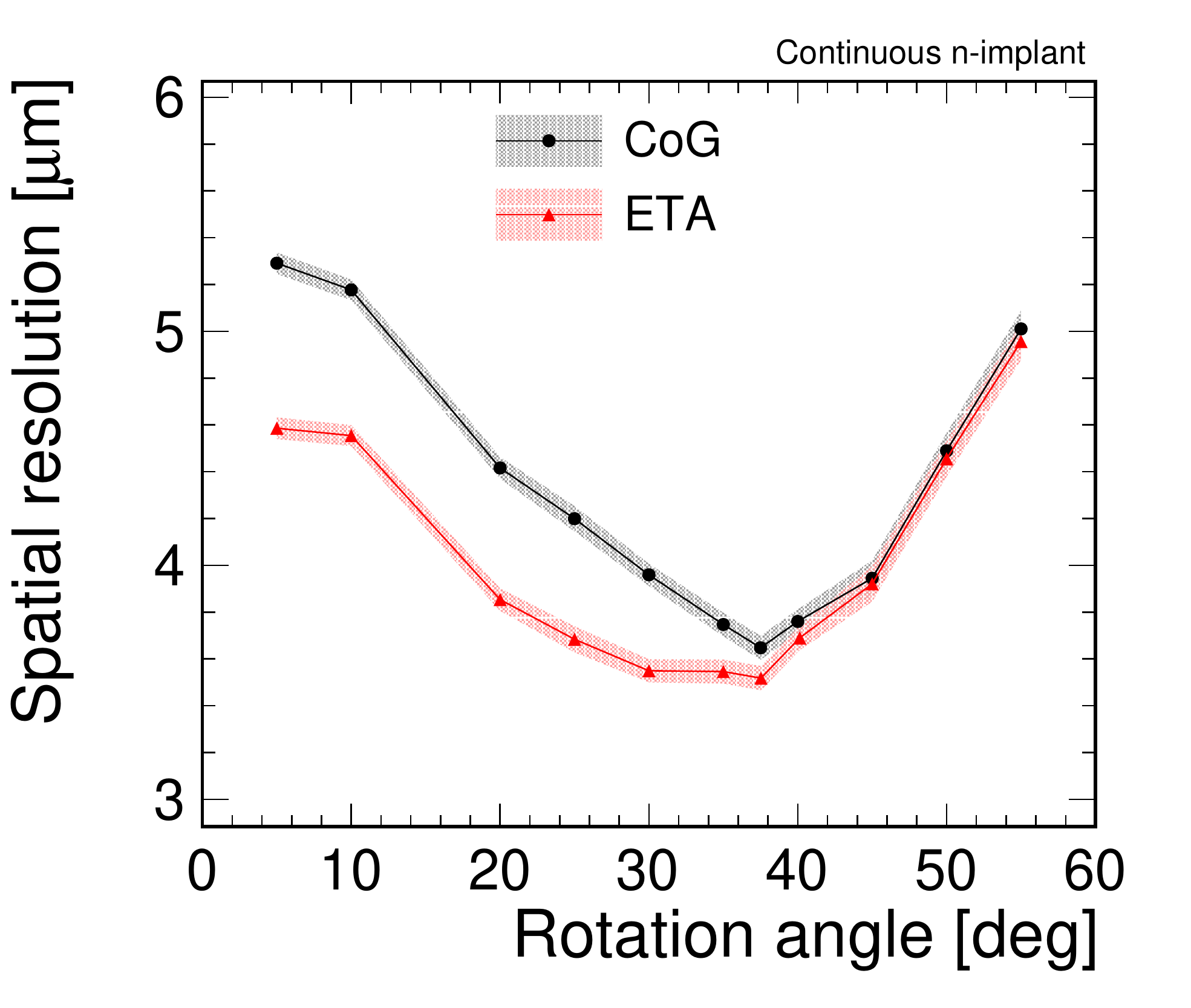}%
	\caption{Spatial resolution as a function of the rotation angle using a charge-weighted centre-of-gravity algorithm (CoG) and an $\eta$-correction (ETA) to reconstruct the cluster position on the DUT.
	A bias voltage of -6\,V/-6\,V was applied to the p-well/substrate.}
	\label{fig:spatialResVsAngle_202110}
\end{figure}

The spatial resolution in row direction improves with increasing rotation angle until approximately $40^\circ$, where it evaluates to $3.6 \pm \SI{0.2}{\micro m}$ after $\eta$-correction, as illustrated in Fig.~\ref{fig:spatialResVsAngle_202110}.
The $\eta$-correction allows for an improvement in spatial resolution for rotation angles below $40^\circ$.
At higher angles, an increase of cluster size $\geq 3$ complicates the application of the reconstruction algorithms and no improvement with respect to the centre-of-gravity algorithm is achievable. 

\begin{table}[bpt]
	\centering
	\caption{Cluster size (CS) for different rotation angles (RA) using a sensor with epitaxial layer and continuous n-implant operated at a threshold of approximately 150\,e.}
	\label{tab:inclined_sensor_performance}
	\begin{tabular}{ccc}
		\hline
		\toprule
		\textbf{RA [$^\circ$]} & \textbf{CS (row)} & \textbf{CS (col.)} \\ 
		\midrule
		0 & $1.46 \pm 0.01$ & $1.38 \pm 0.01$  \\ 
		50 & $2.19 \pm 0.01$ & $1.41 \pm 0.01$ \\ 
		70 & $3.78 \pm 0.01$ & $1.46 \pm 0.01$  \\ 
		\bottomrule
	\end{tabular}
\end{table}

\subsection{Determination of Active Sensor Depth}

The extent of the active sensor volume is an essential ingredient to maximise the signal and thus optimise the sensor performance. 
The results from the previous sections imply that the active sensor volume only covers the upper part of the sensors with epitaxial layer, since thinning the devices down to \SI{50}{\micro \meter} has no significant impact on the performance.   

To quantify the thickness of the active sensor volume, grazing angle measurements~\cite{meroli2011grazing} are performed, whereby inclined particle tracks are used to determine an equivalent charge-collection depth for the observed cluster size.

\begin{figure*}[tbp]
	\centering
	\includegraphics[width=0.55\textwidth]{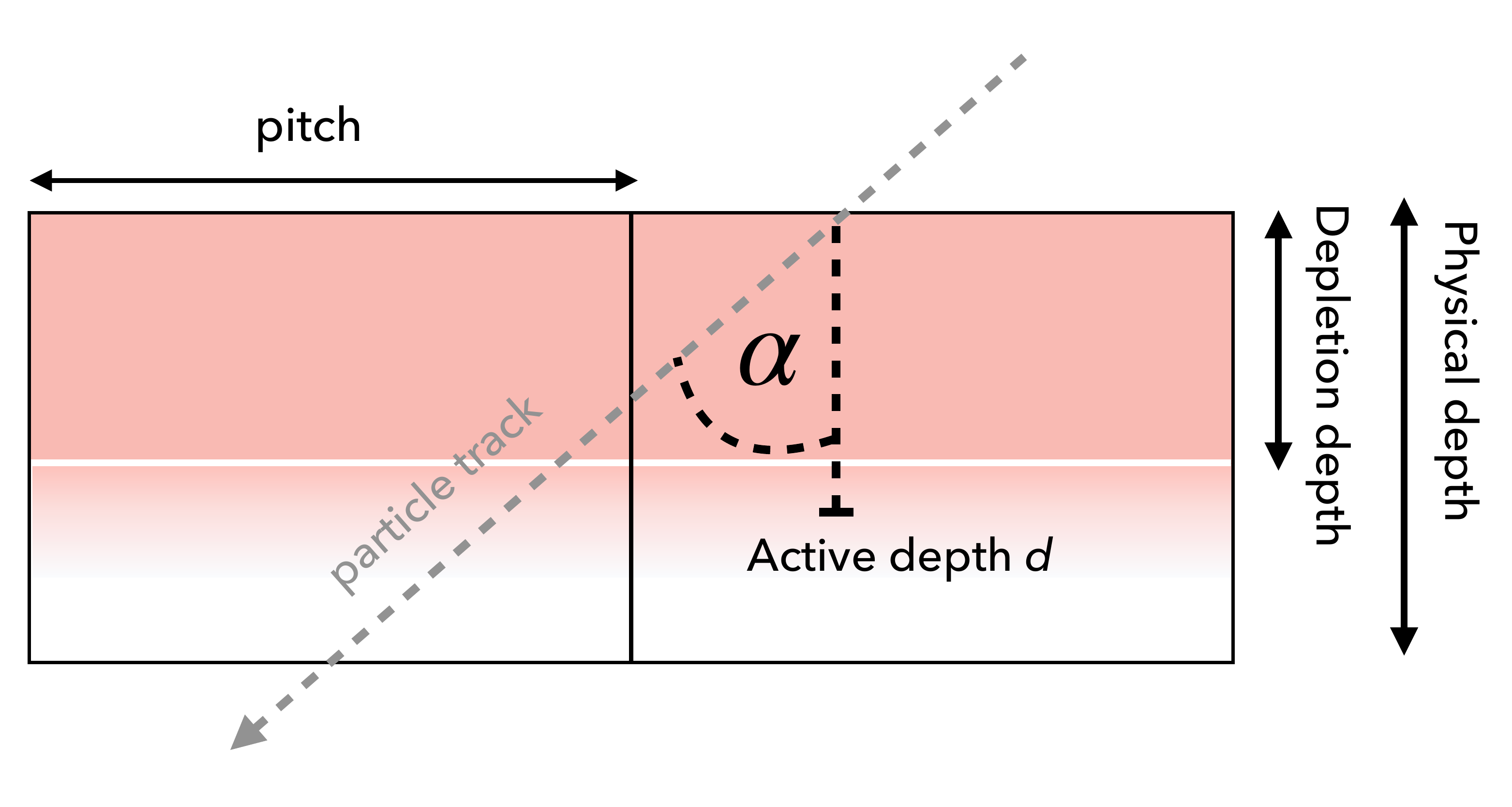}%
	\caption{Schematic representation of the cluster size dependence on the inclination angle of the particle track.}
	\label{fig:active_depth_estimation_sketch}
\end{figure*}

\begin{figure}[bpt]
	\centering
	\includegraphics[width=\columnwidth]{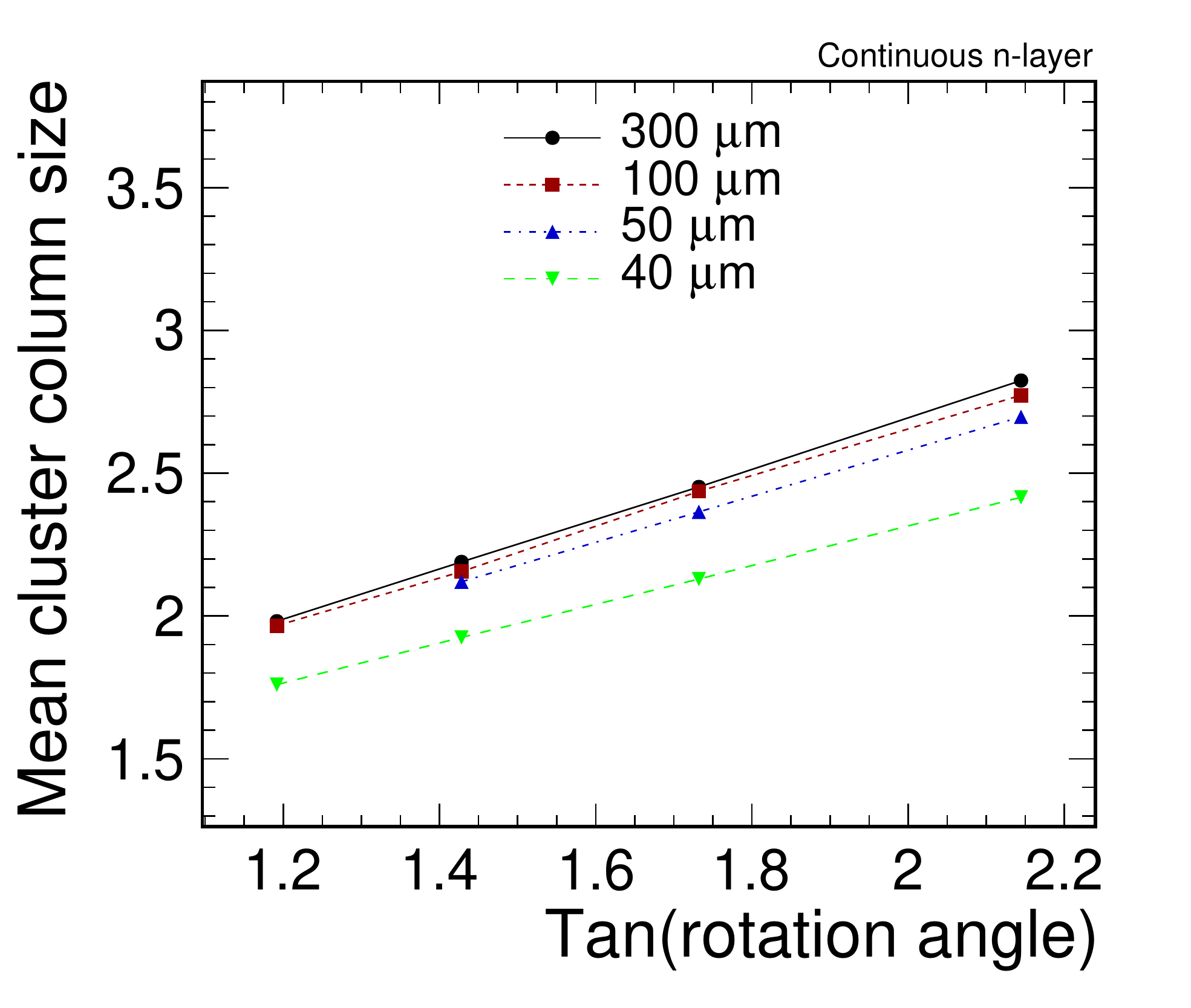}%
	\caption{Mean cluster size in column direction as a function of the tangent of the rotation angle for a sensor with epitaxial layer and continuous n-implant.
	A bias voltage of -6\,V/-6\,V was applied to the p-well/substrate.}
	\label{fig:activeDepth_thickness}
\end{figure}

The estimation of the active sensor depth is based on geometrical consideration, as sketched in Fig.~\ref{fig:active_depth_estimation_sketch}.
The model relates the cluster size in the tilt direction to the incident angle $\alpha$ and the active depth $d$. 
Charge carriers created below the active depth are assumed to have no effect on the cluster size.
The following geometrical relation is considered to extract the active depth $d$ for a sensor tilted in column direction:
\begin{equation}
	\textrm{column cluster size} = \frac{d \tan{\alpha}}{\textrm{pitch}} + s_0,
	\label{eq:active_depth}
\end{equation}
where $s_0$ is the cluster size in column direction for no rotation ($\alpha = 0$). 
The active depth is extracted with a linear fit to the mean cluster size as a function of the tangent of the rotation angle, as exemplified in Fig.~\ref{fig:activeDepth_thickness} for the pixel flavour with continuous n-implant using sensors with different sensor thicknesses. 
The model neglects charge sharing that is not induced by rotation, i.e. charge sharing via diffusion is not accounted for. 
Since the cluster size at small rotation angles is dominated by diffusion effects, data points below $40^\circ$ are excluded from the fit~\cite{dort2021clictd}. 
The effect of diffusion-induced charge sharing is considered in the systematic uncertainties by repeating the fit with varied fit ranges.

The fit results for both pixel flavours are summarised in Table~\ref{tab:active_depth_tb}.
For all sensor types, the estimated active depth of about \SI{30}{\micro \meter} is larger than the depletion depth of $21 \pm \SI{1}{\micro \meter}$ expected from simulation studies~\cite{ballabriga2022transient}.
A non-negligible contribution of charge carriers from the undepleted region is possible, since there is still a residual electric field below the depletion line.

The estimated active depth agrees well with the nominal thickness of the epitaxial layer, which indicates that charge carriers from the undepleted low-resistivity substrate are negligible due to their small lifetime.
Only the active depth for the \SI{40}{\micro \meter} sensor is clearly smaller compared to the other sensors, which is in agreement with the results from the previous sections, where the reduced signal was attributed to the removal of active material.

\begin{table}[bpt]
	\centering
	\caption{Active depth ($d$) for both pixel flavours (Fl.) and different sensor thicknesses (Thickn.).}
	\label{tab:active_depth_tb}
	\begin{tabular}{ccc}
		\toprule
		\textbf{Fl.} & \textbf{Thickn. [$\SI{}{\micro \meter}$]} & \textbf{$d$ [$\SI{}{\micro \meter}$]} \\ 
		\midrule 
		C & 300 & $31.4 \pm 0.1 \textrm{ (stat.)} ^{+ 0.2}_{- 2.4}  \textrm{ (syst.)}$ \\
		C & 100 &$30.7 \pm 0.1 \textrm{ (stat.)} ^{+ 0.3}_{- 1.8}  \textrm{ (syst.)}$  \\
		C & 50 & $29.4 \pm 0.1 \textrm{ (stat.)} ^{+ 0.9}_{- 1.0}  \textrm{ (syst.)}$  \\
		C& 40 & $26.2 \pm 0.1 \textrm{ (stat.)} ^{+ 0.8}_{- 1.0}  \textrm{ (syst.)}$  \\
		\midrule
		S& 300 & $30.8 \pm 0.2 \textrm{ (stat.)} ^{+ 0.4}_{- 1.2}  \textrm{ (syst.)}$   \\
		S & 50 & $29.8 \pm 0.1 \textrm{ (stat.)} ^{+ 0.6}_{- 1.0}  \textrm{ (syst.)}$   \\
		\bottomrule
	\end{tabular}
\end{table}

It can be concluded that the CLICTD sensors with an epitaxial layer of \SI{30}{\micro \meter}  can be thinned down to a total thickness of \SI{50}{\micro \meter} without suffering from a significant loss in sensor performance. 
Assuming that a MIP generates on average about 65 -- 80 electron-hole pairs per micrometer~\cite{meroli2011energy}, the expected signal evaluates to about 2000 -- 2400\,e for an active depth of approximately \SI{30}{\micro \meter}. 
For thinner sensors, performance degradations emerge due to the removal or damage of the active sensor volume.

\begin{figure}[tbp]
	\centering
	\includegraphics[width=\columnwidth]{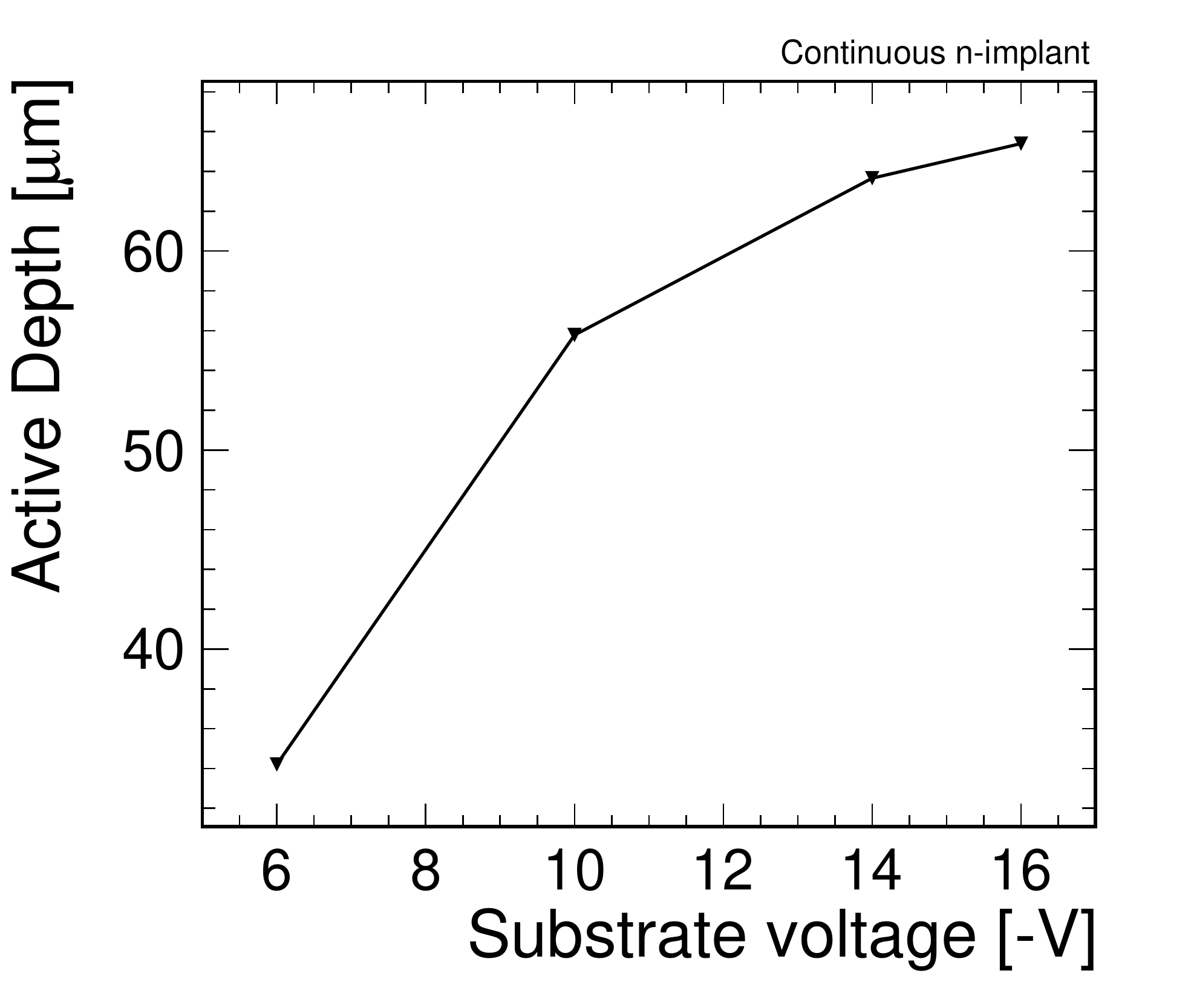}%
	\caption{Active depth as a function of the substrate bias voltage for a sensor with Czochralski substrate with segmented n-implant.}
	\label{fig:activeDepthVsVoltage}
\end{figure}

Unlike for sensors with epitaxial layer, the depletion for the Czochralski substrate is not limited in depth by the thickness of the epitaxial layer. 
The increased depletion region gives access to a larger active sensor volume, as illustrated in the measurements shown in Fig.~\ref{fig:activeDepthVsVoltage}, where the active depth as a function of the substrate voltage is depicted for a Czochralski sensor with segmented n-implant.

The active depth at a substrate voltage of -6\,V evaluates to 
$$34.2 \pm 0.1 \textrm{ (stat.)} ^{+ 1.5}_{- 0.6}  \textrm{ (syst.)}$$
and is therefore slightly larger compared to the sensors with epitaxial layer. 
With higher absolute substrate voltages, the active depth increases and reaches  $$65.4 \pm 0.1 \textrm{ (stat.)} ^{+ 0.5}_{- 0.7}  \textrm{ (syst.)}$$
at a substrate voltage of -16\,V.
At this voltage, the active depth is more than twice as large as the depth for the sensors with epitaxial layer resulting in a significant increase in signal, which is expected to be around 4200 -- 5200\,e.
The higher signal translates into a better performance as shown in the previous section. 

However, the improvement is limited by the front-end, which is not optimised for the large signal generated in the thick Czochralski substrate.

\section{Summary \& Outlook}

The performance, charge-sharing properties and the active sensor depth were investigated for the small collection-electrode monolithic CMOS sensor CLICTD.
Different thicknesses for samples with a thin epitaxial layer were studied and the performance was found to be similar for sensors between \SI{50}{\micro \meter} and \SI{300}{\micro \meter}.
Sensors thinned down to \SI{40}{\micro \meter} exhibited a degradation in performance, which was attributed to a smaller active sensor depth as determined by grazing angle measurements. 
The active depth of the thicker sensors was found to correspond to the nominal thickness of \SI{30}{\micro \meter} of the epitaxial layer. 

To achieve a larger active depth and thus a higher signal, CLICTD sensors fabricated on \SI{100}{\micro \meter} thick Czochralski substrate were tested and a twofold increase in active depth was found using a substrate bias voltage of -16\,V. 
The total signal is expected to double as well and is shared among more pixel cells.
As a consequence, an improvement of approximately \SI{15}{\percent} in spatial and \SI{14}{\percent} in time resolution was determined in combination with an improved efficiency at high detection thresholds.
The design of the front-end would have to be modified in order to exhaust the full  potential of the Czochralski substrate.
 
The sensor performance was also evaluated for inclined particle tracks and an improved performance was found due to the longer particle path through the active sensor volume resulting in a higher signal. 
The spatial resolution has an optimum at an inclination angle of $40^\circ$, where it evaluates to $3.6 \pm \SI{0.2}{\micro \meter}$ after $\eta$-correction.


\section*{Acknowledgements}
\label{sec:acknowledgements}
The authors would like to thank the STREAM network for providing the high-resistivity wafer material and coordinating the sensor production. 

The measurements leading to these results have been performed at the Test Beam Facility at DESY Hamburg (Germany), a member of the Helmholtz Association (HGF).
This project has received funding from the European Union’s Horizon 2020 research and innovation programme under grant agreement No 654168 and under the Marie Skłodowska-Curie grant agreement No 675587 STREAM. 
This work has been sponsored by the Wolfgang Gentner Programme of the German Federal Ministry of Education and Research (grant no. 05E15CHA).
This work was carried out in the framework of the CLICdp Collaboration.

\section*{CRediT authorship statement}

\textbf{R.~Ballabriga} Resources
\textbf{J.~Braach} Investigation
\textbf{E.~Buschmann} Investigation
\textbf{M.~Campbell} Methodology, Resources
\textbf{D.~Dannheim} Investigation, Methodology, Supervision, Writing - Review \& Editing
\textbf{K.~Dort} Formal analysis, Investigation, Software, Visualization, Writing - Original Draft
\textbf{L. Huth} Investigation, Software
\textbf{I.~Kremastiotis} Investigation, Resources
\textbf{J.~Kr\"oger} Investigation, Software
\textbf{L.~Linssen} Project administration, Funding acquisition
\textbf{M.~Munker} Investigation, Methodology, Supervision
\textbf{W.~Snoeys} Conceptualization, Resources
\textbf{S.~Spannagel} Investigation, Software
\textbf{P.~\v{S}vihra} Writing - Review \& Editing
\textbf{T.~Vanat} Investigation, Resources

\bibliography{bibliography}

\end{document}